\documentclass[final,3p,11pt,authoryear]{elsarticle}

\makeatletter
\def\ps@pprintTitle{%
 \let\@oddhead\@empty
 \let\@evenhead\@empty
 \def\@oddfoot{\centerline{\thepage}}%
 \let\@evenfoot\@oddfoot}
\makeatother
\usepackage{color}
\usepackage[usenames,dvipsnames]{xcolor}
\usepackage{hyperref}

\usepackage{natbib}
\biboptions{numbers,sort&compress}
\usepackage[nodots]{numcompress}
\usepackage{graphicx}
\usepackage{color}
\usepackage{mathrsfs}
\usepackage{eufrak}
\usepackage{yfonts}
\usepackage[utf8]{inputenc}
\usepackage{mdframed}
\usepackage{array}
\usepackage{float}
\usepackage{epstopdf}
\usepackage{caption}
\usepackage{subcaption}
\usepackage{amsfonts,amssymb,amsbsy,amsmath,mathtools}

\newcolumntype{M}[1]{>{\centering\arraybackslash}m{#1}}
\newcolumntype{N}{@{}m{0pt}@{}}

\journal{International Journal of Non-Linear Mechanics}
\begin{document}

\begin{frontmatter}

\renewcommand{\thefootnote}{\fnsymbol{footnote}}
\title{\textbf{Nonlinear finite element analysis of lattice core sandwich  plates\let\thefootnote\relax\footnote{{\color{Blue}\textbf{Recompiled, unedited accepted manuscript}}. \copyright 2020. Made available under \href{https://creativecommons.org/licenses/by-nc-nd/4.0/}{{\color{Blue}\textbf{\underline{CC-BY-NC-ND 4.0}}}}}}}




\author[add2]{Praneeth Nampally}
\author[add2,add1]{Anssi T. Karttunen\corref{cor1}}
\cortext[cor1]{Corresponding author. anssi.karttunen@iki.fi. \textbf{Cite as}: \textit{Int. J. Nonlin. Mech.} 2020;121:103423 \href{https://doi.org/10.1016/j.ijnonlinmec.2020.103423}{{\color{OliveGreen}\textbf{\underline{doi link}}}}}
\address[add1]{Aalto University, Department of Mechanical Engineering, Espoo, Finland}
\address[add2]{Texas A\&M University, Department of Mechanical Engineering, College Station, TX, USA}
\author[add2]{J.N. Reddy}

\begin{abstract}
A displacement-based, geometrically nonlinear finite element model is developed for lattice core sandwich panels modeled as 2-D equivalent single-layer (ESL), first-order shear deformation theory (FSDT) micropolar plates. The nonlinearity is due to the moderate macrorotations of the plate which are modeled by including the von K\'arm\'an nonlinear strains in the micropolar strain measures. Weak-form Galerkin formulation with linear Lagrange interpolations is used to develop the displacement finite element model. Selective reduced integration is used to eliminate shear locking and membrane locking. The novel finite element model is used to study the nonlinear bending and linear free vibrations of web-core and pyramid core sandwich panels. Clamped and free edge boundary conditions are considered for the first time for the 2-D micropolar ESL-FSDT plate theory. The present 2-D finite element results are in good agreement with the corresponding detailed 3-D FE results for the lattice core sandwich panels. The 2-D element provides computationally cost-effective solutions; in a nonlinear bending example, the number of elements required for the 2-D micropolar plate is of the order $10^3$, whereas for the corresponding 3-D model the order is $10^5$. 
\end{abstract}
\begin{keyword}
Micropolar plates \sep Constitutive modeling \sep Geometric nonlinearity \sep Lattice material \sep Finite element \sep Nonlinear bending \sep Natural Frequencies


\end{keyword}

\end{frontmatter}


\section{Introduction}

The rapid growth of manufacturing technologies has enabled the design and development of materials whose microstructure can be architected to achieve desired functionality, including high stiffness-to-weight ratios \citep{fleck2010}. The scale of the architected microstructure can range from a few nanometers \citep{bauer2017} to several meters. Lattice materials used in sandwich panels are a class of architected materials whose microstructure is typically in the order of centimeters \citep{allen1969,birman2018}. A variety of manufacturing techniques are available for the production of sandwich panels \citep{Karlsson1997,wadley2003}. Conventional sandwich panels consist of two face sheets and a relatively low-stiffness core between them. The face sheets and core can be of same or different materials depending on the functionality required \citep{vinson2001}. Sandwich panels are designed so that the face sheets take the bending loads while the core carries most of the shear loading \citep{johnson1986}. 

Sandwich panels have received a lot of attention because of their superior performance compared to their monolithic counterparts made solely of either the face sheet material or the core material. For example, sandwich panels have found applications in aerospace industry \citep{arunkumar2017,cherniaev2017} and marine industry \citep{bitzer1994,Mouritz2001}. They are also being used in air and underwater blast resistance structures \citep{fatt2017}. The web-core and pyramid core sandwich panels considered in this paper are two regularly considered sandwich constructions \citep{yungwirth2008,xie2017}. Laser-welded web-core steel sandwich panels have found applications in shipbuilding as staircase landings and non-structural walls \citep{kujala2005,roland1997} and also show good potential for applications in bridges and buildings \citep{bright2004,bright2007,nilsson2017,briscoe2011}.

Increase in the applications for lattice core sandwich panels has created a necessity to carry out analytical and computational structural analyses for such panels. Because of the involved geometry of lattice structures, in most cases analytical solutions are very difficult to achieve and even computational methods can quickly get expensive when full geometric details of a lattice core are considered. Hence various techniques have been developed to analyze lattice structures in an effort to cut down the computational costs and achieve a similar level of accuracy as a complete 3-D computational analysis provides. For example, \cite{birman2018}, \cite{carrera2009}, \cite{burton1995} and \cite{sayyad2017} gave reviews on various modeling techniques employed in the study of lattice structures. \cite{vigliotti2014} and \cite{cohen2019} proposed modeling techniques for capturing the nonlinear response of lattice materials. A review of equivalent single-layer theories for sandwich structures was given by \cite{abrate2017}. 

Of the non-classical continuum mechanics theories to model lattice core sandwich panels, we mention the strain gradient theory \citep{khakalo2018,khakalo2020}, the couple stress theory \citep{goncalves2017}, and the micropolar theory \citep{karttunen2018a,chowdhury2019}, which is particularly well-suited for predicting the structural response of bending-dominated lattice panels accurately. This may be attributed to the additional, independent rotational degrees of freedom the micropolar theory provides. Detailed bending-dominated lattice unit cells may be constructed using beam and shell finite elements and the micropolar theory allows us to pass information related to both the translational and the \textit{rotational} degrees of freedom of the beam and shell elements from a detailed FE model into, for example, a 2-D ESL plate model through a homogenization process. Recently \cite{karttunen2019} used the micropolar theory to model 3-D web-core sandwich panels as 2-D orthotropic equivalent-single layer first-order shear deformation (ESL-FSDT) micropolar plates. Linear transverse deflections and free linear vibrations of web-core lattice plate were considered by using the Navier solution for simply-supported plates.

Micropolar plate theories in the literature usually consider only linearized micropolar strains and wryness tensor components \citep{eringen1967} where the microrotations and all the displacement gradients are assumed to be small. While the assumption that the displacement gradients are small is usually valid at low load intensities, the plates may exhibit nonlinear behavior as the load intensity is increased. These nonlinear deflections are often due to the moderate macrorotations the plate undergoes as the load increases. One way to account for the nonlinearity in micropolar plates without compromising on simplicity or accuracy is to revert back to the way nonlinear plate theories based on classical elasticity for moderate macrorotations are constructed. The well-known nonlinear von K\'arm\'an strains have been used in the construction of beam and plate theories based on non-classical continuum theories such as strain gradient theory \citep{ansari2012}, modified couple stress theory \citep{reddy2012} and the constrained Cosserat theory \citep{srinivasa2013}. \cite{ding2016} used von K\'arm\'an nonlinearity to construct nonlinear micropolar beam. More recently, \cite{nampally2019} showed that von K\'arm\'an nonlinearity in micropolar Timoshenko beam theory can accurately predict the structural response of various lattice core sandwich beams. While some nonlinear micropolar plate theories have been proposed, for example, by \cite{ansari2017}, these nonlinear theories do not include the type of constitutive equations that would make them suitable for structures that involve lattice materials. In this paper, we will use the von K\'arm\'an nonlinearity to account for the moderate macrorotations of micropolar plates and the theory is subsequently used in the analysis of lattice core sandwich plates. With this in mind, the aim of the present paper is two-fold: 
\begin{enumerate}
    \item To construct the first finite element model for linear and nonlinear analyses of the 2-D micropolar ESL-FSDT plate theory used to model 3-D sandwich panels
    \item To analyse, for different boundary conditions, the nonlinear bending and linear free vibrations of web-core and pyramid core sandwich panels by using the finite elements.
\end{enumerate}  
There has been a considerable number of finite element formulations for micropolar elasticity in general and micropolar plates in particular. Here we mention a few of the finite element formulations in the literature. \cite{bauer2010} constructed 3-D finite elements for the study of large deformation in micropolar solids. \cite{kavsov2013} presented finite element model for the bending analysis of micropolar elastic plates. \cite{godio2015} proposed a displacement-based finite element model for micropolar plates, \cite{ansari2017} proposed a nonlinear finite element model for micropolar plates where the nonlinear micropolar strains were considered. \cite{ansari2018} used 3-D micropolar elements to study the vibrations of micro-beams and micro-plates. To the authors' knowledge there has not been a geometric nonlinear finite element formulation for micropolar plates which have been enriched with von K\'arm\'an nonlinearity to account for moderate macrorotations in a relatively simple way. In this paper, we will develop a displacement-based weak-form Galerkin finite element model for such plates.

The rest of the paper is organized as follows. In Section 2, a brief review of the micropolar equivalent single-layer first-order shear deformation (ESL-FSDT) plate theory for lattice core sandwich panels is given. Nonlinear strains for moderate macrorotations are defined and corresponding governing equations are derived using Hamilton's principle. In Section 3, the displacement-based weak-form Galerkin finite element model is formulated. The nonlinear iterative procedure is discussed as well as the techniques used to avoid shear and membrane locking and the eigenvalue problem formulation used for linear vibration analysis. Time-dependent problems are not considered in this paper. However, the presented finite element model can be extended to time-dependent cases by employing appropriate time discretization schemes. In Section 4, various nonlinear bending and linear free vibration examples of lattice core sandwich panels are considered. The results from the formulated 2-D plate finite element model are compared with the corresponding 3-D finite element results. Finally, concluding remarks are given in Section 5. 
\section{Micropolar plate model}
Here, we briefly revise the equivalent single-layer first-order shear deformation theory (ESL-FSDT) of micropolar plates presented by \cite{karttunen2019}.
\subsection{Displacements and microrotations}
The 3-D displacements and microrotations of a plate-like micropolar continuum can be approximated by 2-D midsurface kinematic variables ($u_0, v_0, w_0, \phi_x, \phi_y, \psi_x, \psi_y$) so that
\begin{equation}
\begin{aligned}
u_1(x,y,z,t)&=u_0(x,y,t)+z\phi_x(x,y,t) \\
u_2(x,y,z,t)&=v_0(x,y,t)+z\phi_y(x,y,t) \\
u_3(x,y,z,t)&=w_0(x,y,t) \\
\psi_1(x,y,z,t)&=\psi_x(x,y,t) \\
\psi_2(x,y,z,t)&=\psi_y(x,y,t) \\
\psi_3(x,y,z,t)&=0
\end{aligned}
\end{equation}
where $t$ is time, $(u_0, v_0, w_0)$ denote the displacements of a point on the plane $z=0$, and ($\phi_x,\phi_y$) are the rotations of a transverse normal about the $y$- and $x$-axes, respectively, whereas ($\psi_x,\psi_y$) are microrotations about the $x$- and $y$-axes, respectively. Finally, the following two assumptions are introduced here for the first-order shear deformation theory (FSDT): (1) the formulation is for plates of constant thickness, which (2) do not possess a drilling degree of freedom [$\psi_3(x,y,z)=0$].
\subsection{Strains}
In the micropolar theory, the linearized microstrain and wryness tensors are defined as \citep{eringen2012}
\begin{align}
    \varepsilon_{kl} &= u_{l,k} + \epsilon_{lkm}\psi_{m} \\
    \chi_{kl} &= \psi_{l,k}
\end{align}
It follows that for the FSDT plate at hand, the nonzero strains in Cartesian coordinates are
\begin{equation}
 \begin{aligned}
    \varepsilon_{xx} &= \frac{\partial u_{0}}{\partial x} + z\frac{\partial \phi_{x}}{\partial x} =  \varepsilon_{xx}^{0} + z\kappa_{xx}, \qquad
    \varepsilon_{yy} = \frac{\partial v_{0}}{\partial y} + z\frac{\partial \phi_{y}}{\partial y}  =  \varepsilon_{yy}^{0} + z\kappa_{yy} \\
    \varepsilon_{xy} &= \frac{\partial v_{0}}{\partial x} + z\frac{\partial \phi_{y}}{\partial x} =  \varepsilon_{xy}^{0} + z\kappa_{xy},  \qquad
    \varepsilon_{yx} = \frac{\partial u_{0}}{\partial y} + z\frac{\partial \phi_{x}}{\partial y}   =  \varepsilon_{yx}^{0} + z\kappa_{yx} \\
    \varepsilon_{xz} &= \frac{\partial w_{0}}{\partial x} + \psi_{y}, \qquad \qquad \qquad \qquad \quad 
    \varepsilon_{zx} = \phi_{x} - \psi_{y} \\
    \varepsilon_{yz} &= \frac{\partial w_{0}}{\partial y} - \psi_{x}, \qquad \qquad \qquad \qquad \quad 
    \varepsilon_{zy} = \phi_{y} + \psi_{x}  \\
    \chi_{xx} & = \frac{\partial \psi_{x}}{\partial x}, \quad \chi_{yy} = \frac{\partial \psi_{y}}{\partial y}, \qquad \qquad \quad 
    \chi_{xy}  = \frac{\partial \psi_{y}}{\partial x}, \quad \chi_{yx}  = \frac{\partial \psi_{x}}{\partial y}
\end{aligned} 
\end{equation}
With the inclusion of von K\'arm\'an type geometric nonlinearities \citep{reddy2015} into the vector $\boldsymbol{\epsilon}^0$ below, we write the strains in the form
\begin{equation}
    \boldsymbol{\epsilon}^0= 
    \begin{Bmatrix}
     \varepsilon^{0}_{xx} \\
     \varepsilon^{0}_{yy} \\
     \varepsilon^{0}_{xy} \\
     \varepsilon^{0}_{yx} \\
    \end{Bmatrix} =
    \begin{Bmatrix}
     \frac{\partial u_{0}}{\partial x} + \frac{1}{2}\left(\frac{\partial w_{0}}{\partial x}\right)^2  \\
     \frac{\partial v_{0}}{\partial y} + \frac{1}{2}\left(\frac{\partial w_{0}}{\partial y}\right)^2  \\
     \frac{\partial v_{0}}{\partial x} + \frac{1}{2}\frac{\partial w_{0}}{\partial x} \frac{\partial w_{0}}{\partial y} \\
     \frac{\partial u_{0}}{\partial y} + \frac{1}{2} \frac{\partial w_{0}}{\partial x} \frac{\partial w_{0}}{\partial y} \\
    \end{Bmatrix}  \qquad
    \boldsymbol{\kappa}= 
    \begin{Bmatrix}
     \kappa_{xx} \\
     \kappa_{yy} \\
     \kappa_{xy} \\
     \kappa_{yx} \\
    \end{Bmatrix} =
    \begin{Bmatrix}
     \frac{\partial \phi_{x}}{\partial x} \\
     \frac{\partial \phi_{y}}{\partial y} \\
     \frac{\partial \phi_{y}}{\partial x} \\
     \frac{\partial \phi_{x}}{\partial y} \\
    \end{Bmatrix} \\
\end{equation}

\begin{equation}
    \boldsymbol{\gamma}= 
    \begin{Bmatrix}
        \gamma^{s}_{x} \\
        \gamma^{a}_{x} \\
        \gamma^{s}_{y} \\
        \gamma^{a}_{y} \\
    \end{Bmatrix} =
    \begin{Bmatrix}
        \frac{\partial w_{0}}{\partial x} + \phi_{x} \\
        \frac{\partial w_{0}}{\partial x} - \phi_{x} + 2\psi_{y} \\
        \frac{\partial w_{0}}{\partial y} + \phi_{y} \\
        \frac{\partial w_{0}}{\partial y} - \phi_{y} - 2\psi_{x} \\ 
    \end{Bmatrix} \qquad
    \boldsymbol{\chi}=
    \begin{Bmatrix}
        \chi_{xx} \\
        \chi_{yy} \\
        \chi_{xy} \\
        \chi_{yx} \\
    \end{Bmatrix} =
    \begin{Bmatrix}
        \frac{\partial \psi_{x}}{\partial x} \\
        \frac{\partial \psi_{y}}{\partial y} \\
        \frac{\partial \psi_{y}}{\partial x} \\
        \frac{\partial \psi_{x}}{\partial y} \\
    \end{Bmatrix}
\end{equation}
where the symmetric shear strains are defined as 
\begin{equation}
\begin{aligned}
\gamma^s_x&=\epsilon_{xz}+\epsilon_{zx}=\frac{\partial w_0}{\partial x}+\phi_x \\
\gamma^s_y&=\epsilon_{yz}+\epsilon_{zy}=\frac{\partial w_0}{\partial y}+\phi_y
\end{aligned}
\end{equation}
and the antisymmetric shear strains are
\begin{equation}
\begin{aligned}
\gamma^a_x&=\epsilon_{xz}-\epsilon_{zx}=\frac{\partial w_0}{\partial x}-\phi_x+2\psi_y=2(\psi_y-\omega_{2}) \\
 \gamma^a_y&=\epsilon_{yz}-\epsilon_{zy}=\frac{\partial w_0}{\partial y}-\phi_y-2\psi_x=2(\omega_{1}-\psi_x)
\end{aligned}
\end{equation}
where $(\omega_{1},\omega_{2})$ are the macrorotations. The symmetric shear strains $(\gamma^s_x,\gamma^s_y)$ take the same forms as the shear strains in the conventional ESL-FSDT (ESL-Mindlin) plate theory founded on classical elasticity. The antisymmetric parts are defined by the macrorotations and the microrotations.
\subsection{Constitutive equations}
\begin{figure}
\centering
\includegraphics[scale=0.45]{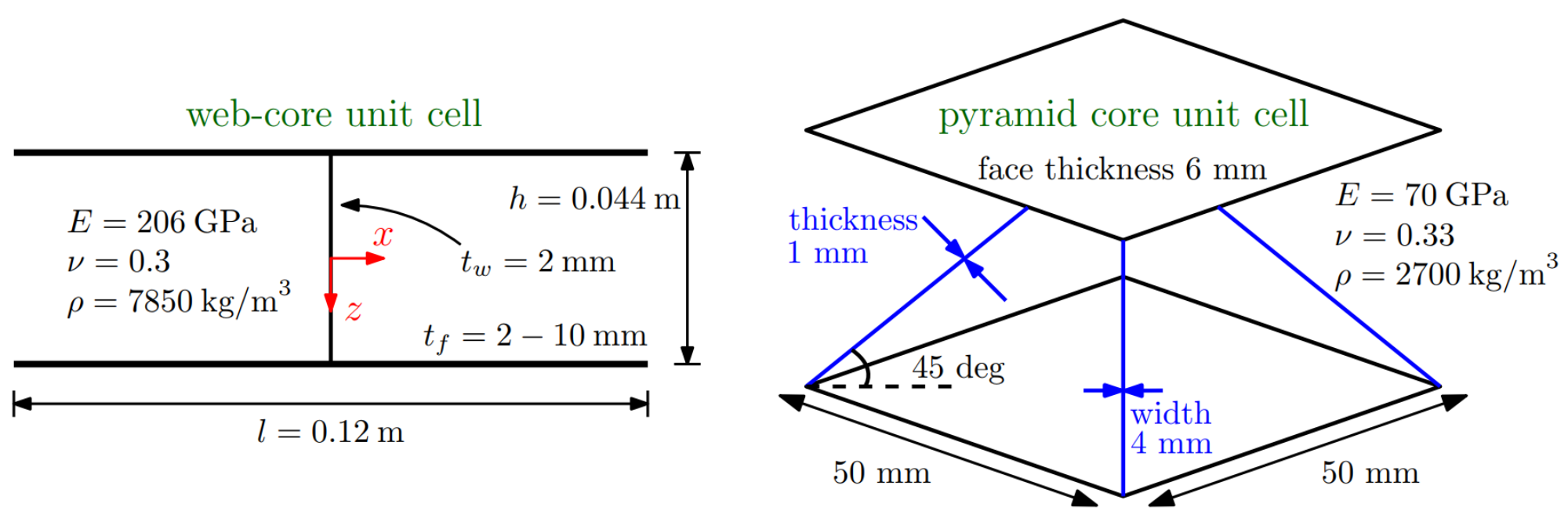}
\caption{Parameters of web-core and pyramid core unit cells made of steel and aluminium, respectively. All face sheet edges of both cores are taken to be of equal length so that, e.g., the web-core planform area is $A=l^2=0.0144$ m$^2$. The struts (beams) in the pyramid core have rectangular cross sections.}
\end{figure}
The unit cells for the web-core and pyramid core are presented in Fig.~1. The unit cells represent \textit{lattice materials} of which the 2-D micropolar ESL-FSDT plate is made of. Two-scale, energy-based constitutive modeling was carried out in detail by \cite{karttunen2019} for the web-core unit cell presented in Fig.~1. The used approach also applies to the pyramid core as such. Therefore, we only briefly review the results of the constitutive modeling here. As the outcome of the constitutive modeling, we have for the 2-D micropolar plate continuum
\begin{equation}
\mathbf{S}=\mathbf{C}\boldsymbol{\epsilon}
\end{equation}
where $\mathbf{S}$ is the stress resultant vector and $\mathbf{C}$ is the constitutive matrix. The explicit matrix form of Eq.~(9) is
\begin{equation}
\begin{Bmatrix}
\mathbf{N} \\
\mathbf{M} \\
\mathbf{Q} \\
\mathbf{P}
\end{Bmatrix}
=
\begin{bmatrix}
 \mathbf{A} & \mathbf{0} & \mathbf{0} & \mathbf{0}  \\
 \mathbf{0} & \mathbf{D} & \mathbf{0} & \mathbf{0}  \\
 \mathbf{0} & \mathbf{0} & \mathbf{G} & \mathbf{0}  \\
 \mathbf{0} & \mathbf{0} & \mathbf{0} & \mathbf{H}  \\
\end{bmatrix}
\begin{Bmatrix}
\boldsymbol{\epsilon}^0 \\
\boldsymbol{\kappa} \\
\boldsymbol{\gamma} \\
\boldsymbol{\chi}
\end{Bmatrix}
\end{equation}
where the vectors for the membrane $\mathbf{N}$, global bending and twisting $\mathbf{M}$, symmetric and antisymmetric shear $\mathbf{Q}$ and local (couple-stress related) bending and twisting $\mathbf{P}$ resultants read
\begin{equation}
\begin{aligned}
\mathbf{N}&=\left\{N_{xx} \ \ N_{yy} \ \ N_{xy} \ \ N_{yx} \right\}^{\textrm{T}} \\
\mathbf{M}&=\left\{M_{xx} \ \ M_{yy} \ \ M_{xy} \ \ M_{yx} \right\}^{\textrm{T}} \\
\mathbf{Q}&=\left\{Q^s_{x} \ \ Q^a_{x} \ \ Q^s_{y} \ \ Q^a_{y} \right\}^{\textrm{T}} \\
\mathbf{P}&=\left\{P_{xx} \ \ P_{yy} \ \ P_{xy} \ \ P_{yx} \right\}^{\textrm{T}}
\end{aligned}
\end{equation}
respectively. The submatrices for the constitutive parameters of the web-core and pyramid core in this study are
\begin{align}
&\mathbf{A}=
\begin{bmatrix}
 A_{11} & A_{12} & 0 & 0  \\
 A_{12} & A_{22} & 0 & 0  \\
 0 & 0 & A_{33} & A_{34}  \\
 0 & 0 & A_{34} & A_{44}  \\
\end{bmatrix},
\quad
\mathbf{D}=
\begin{bmatrix}
 D_{11} & D_{12} & 0 & 0  \\
 D_{12} & D_{22} & 0 & 0  \\
 0 & 0 & D_{33} & D_{34}  \\
 0 & 0 & D_{34} & D_{44}  \\
\end{bmatrix} \\
&\mathbf{G}=
\begin{bmatrix}
 G_{11} & G_{12} & 0 & 0  \\
 G_{12} & G_{22} & 0 & 0  \\
 0 & 0 & G_{33} & G_{34}  \\
 0 & 0 & G_{34} & G_{44}  \\
\end{bmatrix},
\quad
\mathbf{H}=
\begin{bmatrix}
 H_{11} & H_{12} & 0 & 0  \\
 H_{12} & H_{22} & 0 & 0  \\
 0 & 0 & H_{33} & H_{34}  \\
 0 & 0 & H_{34} & H_{44}  \\
\end{bmatrix}
\end{align}
The matrices include 24 constitutive parameters and are symmetric in all cases in this study. The novel parameter values obtained for the pyramid core from the constitutive modeling process presented by \cite{karttunen2019} are given in Appendix B.
\subsection{Variational formulation of equations of motion}
The strain energy for the 2-D micropolar plate takes the form 
\begin{equation}
U=\frac{1}{2}\int_\Omega \boldsymbol{\epsilon}^{\textrm{T}}\mathbf{C}\boldsymbol{\epsilon}\ dxdy
\end{equation}
The total kinetic energy of the plate is 
\begin{equation}
K=\frac{1}{2}\int_{\Omega}\mathbf{\dot{u}}^{\textrm{T}}\mathbf{M}\mathbf{\dot{u}}\ dxdy
\end{equation}
where
\begin{equation}
\mathbf{u}=\left\{u_0 \ \ v_0 \ \ w_0 \ \ \phi_x \ \ \psi_y \ \ \phi_y \ \ \psi_x \right\}^{\textrm{T}}
\end{equation}
and for the web-core and pyramid core we have \citep{karttunen2019}
\begin{equation}
\mathbf{M}=
\begin{bmatrix}
 m_{11} & 0 & 0 & 0 & 0 & 0 & 0 \\
 0 & m_{22} & 0 & 0 & 0 & 0 & 0 \\
 0 & 0 & m_{33} & 0 & 0 & 0 & 0 \\
 0 & 0 & 0 & m_{44} & m_{45} & 0 & 0 \\
 0 & 0 & 0 & m_{45} & m_{55} & 0 & 0 \\
 0 & 0 & 0 & 0 & 0 & m_{66} & m_{67} \\
 0 & 0 & 0 & 0 & 0 & m_{67} & m_{77} \\
\end{bmatrix}
\end{equation}
The mass inertia coefficients of Eq.~(17) for the pyramid core at hand are given in Appendix B. For the web-core the coefficients were given in \citep{karttunen2019}. The potential energy contribution due to a distributed transverse load is
\begin{equation}
V=-\int_{\Omega}qw_0\ dxdy
\end{equation}
By substituting expressions (14), (15) and (18) into Hamilton's principle \citep{reddy2019}, we have
\begin{equation}
\delta\int_0^T\left[K-(U+V)\right]dt=0
\end{equation}
which we can write in the form
\begin{equation}
\int_0^T\int_{\Omega}\left(\delta\mathbf{\dot{u}}^{\textrm{T}}\mathbf{M}\mathbf{\dot{u}}-\delta\boldsymbol{\epsilon}^{\textrm{T}}\mathbf{C}\boldsymbol{\epsilon}+q\delta w_0\right)dxdydt=0
\end{equation}
where we can use Eq.~(9), that is, $\mathbf{S}=\mathbf{C}\boldsymbol{\epsilon}$. We arrive at the following equations of motion (Euler–Lagrange equations) of the 2-D micropolar plate essentially by applying integration by parts in Eq.~(20)
\begin{align}
    & \delta u_{0}:\quad \frac{\partial N_{xx}}{\partial x} + \frac{\partial N_{yx}}{\partial y} = m_{11}\frac{\partial^2 u_{0}}{\partial t^2} \\
    & \delta v_{0}:\quad \frac{\partial N_{xy}}{\partial x} + \frac{\partial N_{yy}}{\partial y} = m_{22}\frac{\partial^2 v_{0}}{\partial t^2} \\
    & \delta w_{0}:\quad \frac{\partial( Q^{s}_{x}+ Q^{a}_{x})}{\partial x} + \frac{\partial ( Q^{s}_{y}+ Q^{a}_{y})}{\partial y} + \mathscr{N} + q_{0}= m_{33}\frac{\partial^2 w_{0}}{\partial t^2} \\
    &\delta \phi_{x}: \quad \frac{\partial M_{xx}}{\partial x} + \frac{\partial M_{yx}}{\partial y} - Q^{s}_{x} + Q^{a}_{x} = m_{44}\frac{\partial^2\phi_{x}}{\partial t^2} + m_{45}\frac{\partial^2\psi_{y}}{\partial t^2} \\
    & \delta \psi_{y}: \quad \frac{\partial P_{xy}}{\partial x} + \frac{\partial P_{yy}}{\partial y} - 2Q^{a}_{x} = m_{55}\frac{\partial^2 \psi_{y}}{\partial t^2} + m_{45}\frac{\partial^2 \phi_{x}}{\partial t^2} \\
    &\delta \phi_{y}: \quad \frac{\partial M_{xy}}{\partial x} + \frac{\partial M_{yy}}{\partial y} - Q^{s}_{y} + Q^{a}_{y} = m_{66}\frac{\partial^2\phi_{y}}{\partial t^2} + m_{67}\frac{\partial^2\psi_{x}}{\partial t^2} \\
    & \delta \psi_{x}: \quad \frac{\partial P_{xx}}{\partial x} + \frac{\partial P_{yx}}{\partial y} + 2Q^{a}_{y} = m_{77}\frac{\partial^2 \psi_{x}}{\partial t^2} + m_{67}\frac{\partial^2 \phi_{y}}{\partial t^2} 
\end{align}
where
\begin{align}
    \mathscr{N} = \frac{\partial}{\partial x}\left[N_{xx}\frac{\partial
    w_0}{\partial x} + \frac{1}{2}\left(N_{xy}+N_{yx}\right)\frac{\partial w_{0}}{\partial y}\right] + \frac{\partial}{\partial y}\left[N_{yy}\frac{\partial
    w_0}{\partial y} + \frac{1}{2}\left(N_{xy}+N_{yx}\right)\frac{\partial w_{0}}{\partial x}\right]
\end{align}

\section{Geometrically nonlinear micropolar plate finite element}
\subsection{Finite element formulation}
In this section, we develop the weak form Galerkin finite element model for the governing equations (21)-(27) of the 2-D micropolar ESL-FSDT plate. In the finite element formulation the mid-surface kinematic variables ($u_0, v_0, w_0, \phi_x, \phi_y, \psi_x, \psi_y$) are the primary variables. These variables are approximated using linear Lagrange interpolation functions $L_{j}^{(J)}$ \citep{reddy2019}, where $(J=1,2,3,4,5,6,7)$. Since we are using weak-form Galerkin finite element formulation, the weight functions $w_{i}$ $(i=1,2,3,4,5,6,7)$ are taken to be the same as the Lagrange interpolation functions used in approximating the primary variables. Thus, we have, 
\begin{equation}
\begin{aligned}
    u_{0} &\approx \sum\limits_{j=1}^{4} U_{j}(t)L^{(1)}_{j}(x,y),\qquad w_{1}(x,y) = L^{(1)}_{i}(x,y) \\
    v_{0} &\approx \sum\limits_{j=1}^{4} V_{j}(t)L^{(2)}_{j}(x,y),\qquad w_{2}(x,y) = L^{(2)}_{i}(x,y)\\
    w_{0} &= \sum\limits_{j=1}^{4} W_{j}(t)L^{(3)}_{j}(x,y),\qquad w_{3}(x,y) = L^{(3)}_{i}(x,y)\\
    \phi_{x} &\approx \sum\limits_{j=1}^{4} \Phi x_{j}(t)L^{(4)}_{j}(x,y),\qquad w_{4}(x,y) = L^{(4)}_{i}(x,y)\\
    \psi_{y} &\approx \sum\limits_{j=1}^{4} \Psi y_{j}(t)L^{(5)}_{j}(x,y),\qquad w_{5}(x,y) = L^{(5)}_{i}(x,y)\\
    \phi_{y} &\approx \sum\limits_{j=1}^{4} \Phi y_{j}(t)L^{(6)}_{j}(x,y),\qquad w_{6}(x,y) = L^{(6)}_{i}(x,y)\\
    \psi_{x} &\approx \sum\limits_{j=1}^{4} \Psi x_{j}(t)L^{(7)}_{j}(x,y),\qquad w_{7}(x,y) = L^{(7)}_{i}(x,y)
\end{aligned}
\end{equation}
Now we write the weak form equations of the micropolar plate governing equations (21)-(27) on a typical element $\Omega_{e}$ as 
\begin{align}
     0 =&\int\limits_{\Omega_{e}}\Biggl\{ w_{1}m_{11}\frac{\partial^2 u_{0}}{\partial t^2} + \frac{\partial w_{1}}{\partial x}N_{xx} + \frac{\partial w_{1}}{\partial y}N_{yx} \Biggr\} dxdy - \int\limits_{\tau_{e}}w_{1}Q_{1}ds \\
    0 =& \int\limits_{\Omega_{e}}\Biggl\{w_{2} m_{22}\frac{\partial^2 v_{0}}{\partial t^2}+\frac{\partial w_{2} }{\partial x}N_{xy} + \frac{\partial w_{2}}{\partial y}N_{yy} \Biggr\} dxdy - \int\limits_{\tau_{e}}w_{2}Q_{2}ds 
\end{align}
\begin{align}
    0 =& \int\limits_{\Omega_{e}}\Biggl\{w_{3}m_{33}\frac{\partial^2 w_{0}}{\partial t^2}+ \frac{\partial w_{3}}{\partial x}\left(Q_{x}^{s}+Q_{x}^{a}\right)  + \frac{\partial w_{3}}{\partial y}\left(Q_{y}^{s}+Q_{y}^{a}\right) \nonumber \\
    &\quad + \frac{\partial w_{3}}{\partial x}\left[N_{xx}\frac{\partial
    w_0}{\partial x}  + \frac{1}{2}\left(N_{xy}+N_{yx}\right)\frac{\partial w_{0}}{\partial y}\right]  + \frac{\partial w_{3}}{\partial y}\left[N_{yy}\frac{\partial
    w_0}{\partial y}  + \frac{1}{2}\left(N_{xy}+N_{yx}\right)\frac{\partial w_{0}}{\partial x}\right] \nonumber \\
    &\quad - w_{3}q_{0} \Biggr\} dxdy - \int\limits_{\tau_{e}}w_{3}Q_{3}ds \\
     0 = &\int\limits_{\Omega_{e}}\Biggl\{w_{4}m_{44}\frac{\partial^2\phi_{x}}{\partial t^2} + w_{4}m_{45}\frac{\partial^2\psi_{y}}{\partial t^2}+\frac{\partial w_{4}}{\partial x}M_{xx}  + \frac{\partial w_{4}}{\partial y}M_{yx}  + w_{4}(Q_{x}^{s}-Q_{x}^{a})\Biggr\} dxdy - \int\limits_{\tau_{e}}w_{4}Q_{4}ds  
\end{align}
\begin{align}    
    0 =  &\int\limits_{\Omega_{e}}\Biggl\{w_{5}m_{55}\frac{\partial^2 \psi_{y}}{\partial t^2} + w_{5}m_{45}\frac{\partial^2 \phi_{x}}{\partial t^2} + \frac{\partial w_{5}}{\partial x}P_{xy} + \frac{\partial w_{5}}{\partial y}P_{yy}  + 2w_{5}Q_{x}^{a}\Biggr\} dxdy - \int\limits_{\tau_{e}}w_{5}Q_{5}ds  \\  
    0 = &\int\limits_{\Omega_{e}}\Biggl\{w_{6}m_{66}\frac{\partial^2\phi_{y}}{\partial t^2} + w_{6}m_{67}\frac{\partial^2\psi_{x}}{\partial t^2} + \frac{\partial w_{6}}{\partial x}M_{xy} + \frac{\partial w_{6}}{\partial y}M_{yy} + w_{6}(Q_{y}^{s}-Q_{y}^{a})\Biggr\} dxdy - \int\limits_{\tau_{e}}w_{6}Q_{6}ds \\  
    0 =  & \int\limits_{\Omega_{e}}\Biggl\{w_{7}m_{77}\frac{\partial^2 \psi_{x}}{\partial t^2} + w_{7}m_{67}\frac{\partial^2 \phi_{y}}{\partial t^2}+\frac{\partial w_{7}}{\partial x}P_{xx} + \frac{\partial w_{7}}{\partial y}P_{yx}  - 2w_{7}Q_{y}^{a}\Biggr\} dxdy - \int\limits_{\tau_{e}}w_{7}Q_{7}ds  
\end{align}
After using the constitutive equations (12) and (13) along with the interpolations of the primary variables and weight functions (29) in the above equations, we have the following finite element formulation on a typical element:
\begin{align} \label{eq:37}
    \mathbf{M^{(e)}}\mathbf{\ddot{\mathfrak{U}}^{(e)}} + \mathbf{K^{(e)}}\mathbf{\mathfrak{U}^{(e)}} = \mathbf{F^{(e)}} 
\end{align}
where,
\begin{equation}
\mathbf{M^{(e)}} = 
\begin{bmatrix}
    \mathbf{M^{11}}&\mathbf{M^{12}}&\mathbf{M^{13}}&\mathbf{M^{14}}&\mathbf{M^{15}}&\mathbf{M^{16}}&\mathbf{M^{17}} \\
    \mathbf{M^{21}}&\mathbf{M^{22}}&\mathbf{M^{23}}&\mathbf{M^{24}}&\mathbf{M^{25}}&\mathbf{M^{26}}&\mathbf{M^{27}}  \\
    \mathbf{M^{31}}&\mathbf{M^{32}}&\mathbf{M^{33}}&\mathbf{M^{34}}&\mathbf{M^{35}}&\mathbf{M^{36}}&\mathbf{M^{37}}  \\
    \mathbf{M^{41}}&\mathbf{M^{42}}&\mathbf{M^{43}}&\mathbf{M^{44}}&\mathbf{M^{45}}&\mathbf{M^{46}}&\mathbf{M^{47}}  \\
    \mathbf{M^{51}}&\mathbf{M^{52}}&\mathbf{M^{53}}&\mathbf{M^{54}}&\mathbf{M^{55}}&\mathbf{M^{56}}&\mathbf{M^{57}}  \\
     \mathbf{M^{61}}&\mathbf{M^{62}}&\mathbf{M^{63}}&\mathbf{M^{64}}&\mathbf{M^{65}}&\mathbf{M^{66}}&\mathbf{M^{67}}  \\
    \mathbf{M^{71}}&\mathbf{M^{72}}&\mathbf{M^{73}}&\mathbf{M^{74}}&\mathbf{M^{75}}&\mathbf{M^{76}}&\mathbf{M^{77}}  \\
\end{bmatrix}^{(e)} 
\end{equation}
\begin{equation}
\mathbf{K^{(e)}} = 
\begin{bmatrix}
    \mathbf{K^{11}}&\mathbf{K^{12}}&\mathbf{K^{13}}&\mathbf{K^{14}}&\mathbf{K^{15}}&\mathbf{K^{16}}&\mathbf{K^{17}} \\
    \mathbf{K^{21}}&\mathbf{K^{22}}&\mathbf{K^{23}}&\mathbf{K^{24}}&\mathbf{K^{25}}&\mathbf{K^{26}}&\mathbf{K^{27}}  \\
    \mathbf{K^{31}}&\mathbf{K^{32}}&\mathbf{K^{33}}&\mathbf{K^{34}}&\mathbf{K^{35}}&\mathbf{K^{36}}&\mathbf{K^{37}}  \\
    \mathbf{K^{41}}&\mathbf{K^{42}}&\mathbf{K^{43}}&\mathbf{K^{44}}&\mathbf{K^{45}}&\mathbf{K^{46}}&\mathbf{K^{47}}  \\
    \mathbf{K^{51}}&\mathbf{K^{52}}&\mathbf{K^{53}}&\mathbf{K^{54}}&\mathbf{K^{55}}&\mathbf{K^{56}}&\mathbf{K^{57}}  \\
     \mathbf{K^{61}}&\mathbf{K^{62}}&\mathbf{K^{63}}&\mathbf{K^{64}}&\mathbf{K^{65}}&\mathbf{K^{66}}&\mathbf{K^{67}}  \\
    \mathbf{K^{71}}&\mathbf{K^{72}}&\mathbf{K^{73}}&\mathbf{K^{74}}&\mathbf{K^{75}}&\mathbf{K^{76}}&\mathbf{K^{77}}  \\
\end{bmatrix}^{(e)} 
\end{equation}
\begin{equation}
\mathbf{\ddot{\mathfrak{U}}^{(e)}} =
\begin{Bmatrix}
    \mathbf{\ddot{U}} \\
    \mathbf{\ddot{V}} \\
    \mathbf{\ddot{W}} \\
    \boldsymbol{\ddot{\Phi}x} \\
    \boldsymbol{\ddot{\Psi}y} \\
    \boldsymbol{\ddot{\Phi}y} \\
    \boldsymbol{\ddot{\Psi}x} \\
\end{Bmatrix}^{(e)} \qquad
\mathbf{\mathfrak{U}^{(e)}} =
\begin{Bmatrix}
    \mathbf{U} \\
    \mathbf{V} \\
    \mathbf{W} \\
    \boldsymbol{\Phi x} \\
    \boldsymbol{\Psi y} \\
    \boldsymbol{\Phi y} \\
    \boldsymbol{\Psi x} \\
\end{Bmatrix}^{(e)} \qquad
\mathbf{F^{(e)}} = 
\begin{Bmatrix}
    \mathbf{F^1} \\
    \mathbf{F^2} \\
    \mathbf{F^3} \\
    \mathbf{F^4} \\
    \mathbf{F^5} \\
    \mathbf{F^6} \\
    \mathbf{F^7} \\
\end{Bmatrix}^{(e)}
\end{equation}
The non-zero components of the above matrices are given in Appendix A.
\subsection{Solution of nonlinear equations}
Although the nonlinear finite element equations (37) can be used to solve time-dependent cases with appropriate time discretization schemes, in the present study we only consider time-independent nonlinear cases. For the time-independent nonlinear case the finite element equations (37) are solved using Newton's iterative procedure \citep{reddy2015}, by constructing the tangent stiffness of a typical element at the beginning of $r^{th}$ iteration as
\begin{align}
    &\mathbf{T^{(e)}}^{(r)} = \left[\frac{\partial \mathbf{R^{(e)}}}{\partial \mathbf{\mathfrak{U}^{(e)}}}\right]^{(r-1)} 
\end{align}
such that 
\begin{align}
    &\mathbf{T^{(e)}}^{(r)}\Delta \mathbf{\mathfrak{U}^{e}} = - \mathbf{R^{(e)}}^{(r-1)}
\end{align}
where 
\begin{align*}
    &\mathbf{R^{(e)}}  = \mathbf{K^{(e)}}(\mathbf{{\mathfrak{U}}^{(e)}})\mathbf{\mathfrak{U}^{(e)}} - \mathbf{F^{(e)}} \quad  \textnormal{and} \quad \Delta \mathbf{\mathfrak{U}^{(e)}} = \mathbf{\mathfrak{U}^{(e)}}^{(r)} -\mathbf{\mathfrak{U}^{(e)}}^{(r-1)} 
\end{align*}
The explicit expressions of the components of the element tangent stiffness matrix are given in Appendix A. 

After the element equations (41) have been computed, they are assembled according to the nodal connectivity of the mesh to obtain global equations. Boundary conditions are imposed on the global equations and subsequent equations are solved to obtain the global incremental generalized displacement vector $\Delta \mathbf{\mathfrak{U}}$ at the end of $r^{th}$ iteration. The normalized difference between solution vectors from two consecutive iterations, measured with Euclidean norm, is computed at the end of each iteration. If the value computed is less than a preselected tolerance '$tol$' further iterations are terminated and nonlinear convergence is assumed (for all the nonlinear cases considered in the paper $tol = 10^{-3}$)
\begin{align*}
    \sqrt{\frac{\Delta\mathbf{\mathfrak{U}}\cdot\Delta\mathbf{\mathfrak{U}}}{\mathbf{\mathfrak{U}}^{(r)}\cdot\mathbf{\mathfrak{U}}^{(r)}}}\leqslant tol
\end{align*}
Once the nonlinear convergence is attained, the final global generalized displacement vector is obtained using 
\begin{align}
    \mathbf{\mathfrak{U}}^{(r)} =  \Delta \mathbf{\mathfrak{U}} + \mathbf{\mathfrak{U}}^{(r-1)} 
\end{align}
\subsection{Natural Vibration Frequencies}
In the present study we will only consider the natural frequencies of the lattice plates undergoing linear free vibrations. The natural frequencies can be calculated by solving the eigenvalue problem obtained by substituting $\mathbf{\mathfrak{U}}(x,y,t)=\mathbf{\mathfrak{U_{0}}}(x,y)e^{j\lambda t}$ (where $j=\sqrt{-1}$) into the assembled linear global equations (i.e., nonlinear terms in coefficient matrices are ignored) after the imposition of boundary conditions. Here $\mathbf{\mathfrak{U_{0}}}$ is the global mode shape corresponding to the eigenvalue $\lambda^{2}$. Once the eigenvalues are obtained the natural frequencies [Hz] are calculated using 
\begin{align*}
    f_{i} = \frac{\lambda}{2 \pi}
\end{align*}

It should be noted that the number of eigenvalues obtained will be equal to the number of degrees of freedom in the problem. Thus for the convergence of higher mode shapes a finer mesh is required compared to the lower mode shapes.
\subsection{Shear and Membrane Locking}

Since linear Lagrange interpolation functions are used in the approximation of all the primary variables, the elements become excessively stiff in the thin plate limit because of spurious constraints imposed on the bending energy due to this inconsistent interpolation, resulting in a phenomenon known as \textit{shear locking} \citep{hughes1978,reddy2015,reddy2019}. Consider a plate of dimensions $(a \cdot b)$ being modeled by a single rectangular element. Since linear interpolations are used on both $w_{0}$ and $\phi_{x}$, if $(w_{01},w_{02},w_{03},w_{04})$ and $(\phi_{x1},\phi_{x2},\phi_{x3},\phi_{x4})$ are the nodal values of $w_{0}$ and $\phi_{x}$ respectively, we have 
\begin{align*}
    w_{0} &= w_{01}\left(1-\frac{x}{a}\right)\left(1-\frac{y}{b}\right) + w_{02}\frac{x}{a}\left(1-\frac{y}{b}\right) + w_{03}\frac{x}{a}\frac{y}{b} + w_{04}\left(1-\frac{x}{a}\right)\frac{y}{b} \\
    \phi_{x} &= \phi_{x1}\left(1-\frac{x}{a}\right)\left(1-\frac{y}{b}\right) + \phi_{x2}\frac{x}{a}\left(1-\frac{y}{b}\right) + \phi_{x3}\frac{x}{a}\frac{y}{b} + \phi_{x4}\left(1-\frac{x}{a}\right)\frac{y}{b} 
\end{align*}
\begin{align*}
    \gamma_{x}^{s} &=  \left(\frac{w_{02}-w_{01}+a\phi_{x1}}{a}\right)+ \left(\frac{\phi_{x1}-\phi_{x2}+\phi_{x3}-\phi_{x4}}{ab}\right)xy  \\
    &\quad + \left(\frac{\phi_{x2}-\phi_{x1}}{a}\right)x + \left(\frac{w_{01}-w_{02}+w_{03}-w_{04}+a\phi_{x4}-a\phi_{x1}}{ab}\right)y 
\end{align*}
In the thin plate limit $\gamma_{x}^{(s)}$ approaches zero and this only possible when the constant terms and coefficients of $x$, $y$ and $xy$ of $\gamma_{x}^{(s)}$ are all zero. That is,
\begin{align}
     \frac{w_{01}-w_{02}}{a} &= \phi_{x1} \\
     \frac{w_{03}-w_{04}}{a} &= \phi_{x4} \\
    \phi_{x1}= \phi_{x2}, \quad  & \quad \phi_{x3} = \phi_{x4}
\end{align}
 However, Eq.~(46) implies that $\phi_{x}$ is constant with respect to $x$ and this will pose an unnecessary restriction on bending energy which will manifest as shear locking. A similar argument can be extended to antisymmetric shear strains as well. Various remedies have been proposed in the literature to overcome shear locking, see, for example, \cite{hughes1978,zienkiewicz1971} and \cite{bathe1996}. In the present finite element formulation we use selective reduced integration to overcome the shear locking. That is, we will evaluate the stiffness coefficient terms corresponding to symmetric and antisymmetric shear strains using reduced Gauss quadrature rule \citep{reddy2019}.

With the addition of von K\'arm\'an nonlinearity, bending-stretching coupling is introduced into the plate thereby predicting membrane strain even when only bending forces are applied. But in the cases where the membrane (axial) strains are not physically possible in the plate (example, when all the edges are hinge-supported) the theory will still predict membrane strains. This phenomenon is called \textit{membrane locking} \citep{reddy2015}. To overcome this we will use reduced integration on all the nonlinear terms of the element coefficient matrices.  

It is worth mentioning that although linear elements and consequently reduced integration techniques are used in this paper to overcome shear and membrane locking, it is by no means a necessity. The developed finite element equations (\ref{eq:37}) can easily be used in conjunction with higher-order elements which can alleviate locking problems. However, higher-order elements with equally spaced nodes are prone to oscillations near the end points of a standard interval (known as \textit{Runge effect}). This problem can be eliminated by using higher-order elements with nodes located at Gauss-Lobatto-Legendre (GLL) points \citep{karniadakis1999, payette2014}. 
\section{Numerical Results and Discussion}
3-D finite element models for web-core and pyramid core sandwich panels are discussed in Section 4.1. These FE models are built using Abaqus 2019 to provide reference solutions to which the 2-D results can be compared.

In Section 4.2, we first study the convergence of the finite element calculations by considering the linear static bending of a simply-supported web-core sandwich panel under line and uniformly distributed loads. Second, the nonlinear bending of simply-supported web-core panels is investigated for the same loads. Third, we consider the nonlinear bending of web-core panels that have clamped and free edges as well. 

The web-core lattice is bending-dominated, whereas the pyramid core is stretch-dominated, meaning that the struts of the core do not essentially bend but carry only axial loads, that is, they behave as axial rods. It has been shown earlier for lattice core beams that stretch-dominated cores do not exhibit global nonlinear bending but rather go straight from linear bending to local buckling where individual unit cells basically collapse near supports or point loads \citep{nampally2019}. This type of local buckling behavior is not captured by the current plate model, or by any other 2-D ESL-FSDT plate model to the best of our knowledge. In conclusion, in the case of the pyramid core sandwich panels, we focus only the linear natural vibration frequency calculations in this paper. The natural frequencies of both the pyramid core and web-core plate are studied in Section 4.3.
\subsection{Plate dimensions and 3-D FE reference models}
For the web-core plate two different size plates will be considered; the plate planform area is $(a\cdot b)$ m$^2$ and the studied sizes are $(5.4\cdot 3.6)$ m$^2$ for bending and $(1.8\cdot 1.2)$ m$^2$ for natural frequency calculations. The other relevant dimensions were given in Fig.~1. The corresponding 3-D FE reference model for the larger plate consists of 453600 shell elements of type S8R5 and the smaller one contains 141000 shell elements of type S4R. A pyramid core plate of size $(1\cdot 1)$ m$^2$ is considered in the natural frequency calculations. The corresponding 3-D FE reference model consists of 9600 linear beam elements of type B33, 33885 quadrilateral shell elements of type S8R5 and 1002 triangular elements of type STRI65. All the 3-D FE models are convergent.

The 3-D boundary conditions are imposed in a similar manner as in classical simply-supported, clamped and free edge 3-D solid plate problems. For simply-supported edges, for all nodes $i=1,2,\ldots,n$ of the shell elements on edges $x=(-a/2,a/2)$ (see Fig.~2) we use $U_z^i=U_y^i=Rot_x^i=0$ with reference to the global coordinate system. Analogously, for all nodes on edges $y=(-b/2,b/2)$ we use $U_z^i=U_x^i=Rot_y^i=0$. For clamped edges we have $U_x^i=U_y^i=U_z^i=Rot_x^i=Rot_y^i=Rot_z^i=0$. No boundary conditions are set on free edges.
\subsection{Bending analysis}
\begin{figure}[H]
    \centering
    \includegraphics[scale=0.4]{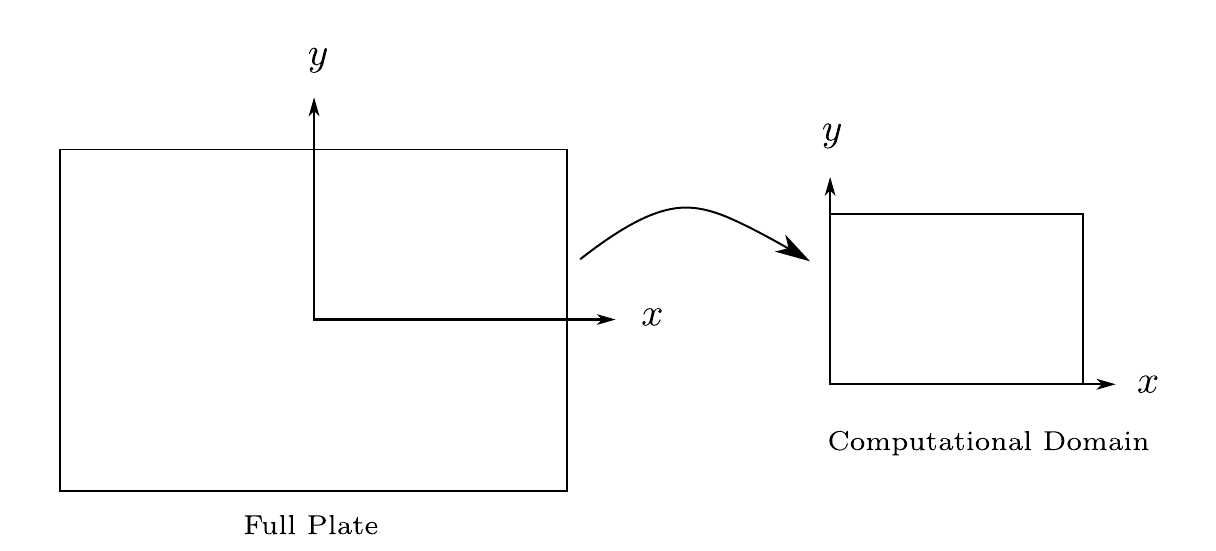}
    \caption{Choice of computational domain for the bending analysis under considered boundary and loading conditions.}
    \label{fig:domain}
\end{figure}

For the bending analysis of the 2-D micropolar ESL-FSDT plates the coordinate system is chosen such that the center of the plate coincides with the origin as shown in  Fig.~\ref{fig:domain}. For the web-core plates the webs are parallel to the $y-$axis. Four sets of boundary conditions are considered:
\begin{enumerate}
    \item Simply-supported on all edges (SSSS). 
    \item Edges parallel to $x$-axis are clamped and edges parallel to $y$-axis are simply supported (CSCS).
    \item All edges are clamped (CCCC). 
    \item Edges parallel to $x$-axis are free and edges parallel to $y$-axis are clamped (CFCF).
\end{enumerate}
Furthermore, for each boundary condition case the plate is subjected to two different loadings, a uniformly distributed load and a line load along the $y$-axis at the center of the plate. Since the boundary and loading conditions considered here result in symmetry about $x$- and $y$-axes, we consider only the quarter plate lying in the first quadrant as the computational domain (see Fig.~\ref{fig:domain}). For such a computational domain the considered boundary conditions after symmetry arguments are listed in Table 1. It should be noted that for web-core plates the center web is along $y$-axis and for the line load case only half of the total load intensity on the full plate is to be considered on the computational domain. Unless stated otherwise all the loads listed in the paper are on full plate.

\begin{table}[H]
\scriptsize
\caption{Various boundary conditions on the computational domain for bending analysis.} 
\centering
  \begin{tabular}{M{0.45in}|M{1.3in}|M{1.3in}|M{1.3in}|M{1.3in} @{}m{0pt}@{}}
    \hline
      &
     SSSS &
     CSCS &
     CCCC &
     CFCF & \\
    \hline
    $y=0$ & $v_{0}=\phi_{y}=\psi_{x}=0$ & $v_{0}=\phi_{y}=\psi_{x}=0$ & $v_{0}=\phi_{y}=\psi_{x}=0$ & $v_{0}=\phi_{y}=\psi_{x}=0$ &\\[20pt]
    \hline
    $y= b/2$ & $u_{0}=w_{0}=\phi_{x}=\psi_{y}=0$ & $ u_{0}=v_{0}=w_{0}=\phi_{x}=\psi_{y}=\phi_{y}=\psi_{x}=0$ & $u_{0}=v_{0}=w_{0}=\phi_{x}=\psi_{y}=\phi_{y}=\psi_{x}=0$ & $N_{yx}=N_{yy}=Q_{y}^{(s)}+Q_{y}^{(a)}=M_{yx}=P_{yy}=M_{yy}=P_{yx}=0$
    &\\[20pt]
    \hline
    $x=0$ & $u_{0}=\phi_{x}=\psi_{y}=0$ & $u_{0}=\phi_{x}=\psi_{y}=0$ & $u_{0}=\phi_{x}=\psi_{y}=0$ & 
    $u_{0}=\phi_{x}=\psi_{y}=0$ & \\[20pt]
    \hline
    $x=a/2$ & $v_{0}=w_{0}=\phi_{y}=\psi_{x}=0$ & $v_{0}=w_{0}=\phi_{y}=\psi_{x}=0$ & $u_{0}=v_{0}=w_{0}=\phi_{x}=\psi_{y}=\phi_{y}=\psi_{x}=0$ &
    $u_{0}=v_{0}=w_{0}=\phi_{x}=\psi_{y}=\phi_{y}=\psi_{x}=0$ & \\[20pt]
    \hline
    $(0,0)$ & $u_{0}=v_{0}=\phi_{x}=\psi_{y}=\phi_{y}=\psi_{x}=0$ & $u_{0}=v_{0}=\phi_{x}=\psi_{y}=\phi_{y}=\psi_{x}=0$ & $u_{0}=v_{0}=\phi_{x}=\psi_{y}=\phi_{y}=\psi_{x}=0$ &
    $u_{0}=v_{0}=\phi_{x}=\psi_{y}=\phi_{y}=\psi_{x}=0$ &  \\[20pt]
    \hline
    $(a/2,0)$ & $v_{0}=w_{0}=\phi_{y}=\psi_{x}=0$ & $ v_{0}=w_{0}=\phi_{y}=\psi_{x}=0$ & $u_{0}=v_{0}=w_{0}=\phi_{x}=\psi_{y}=\phi_{y}=\psi_{x}=0$ &
     $u_{0}=v_{0}=w_{0}=\phi_{x}=\psi_{y}=\phi_{y}=\psi_{x}=0$ &  \\[20pt]
    \hline
    $(0,b/2)$ & $u_{0}=w_{0}=\phi_{x}=\psi_{y}=0$ & $u_{0}=v_{0}=w_{0}=\phi_{x}=\psi_{y}=\phi_{y}=\psi_{x}=0$ & $u_{0}=v_{0}=w_{0}=\phi_{x}=\psi_{y}=\phi_{y}=\psi_{x}=0$ & 
    $u_{0}=\phi_{x}=\psi_{y}=0$ & \\[20pt]
    \hline
    $(a/2,b/2)$ & $u_{0}=v_{0}=w_{0}=\phi_{x}=\psi_{y}=\phi_{y}=\psi_{x}=0$ & $u_{0}=v_{0}=w_{0}=\phi_{x}=\psi_{y}=\phi_{y}=\psi_{x}=0$ & $u_{0}=v_{0}=w_{0}=\phi_{x}=\psi_{y}=\phi_{y}=\psi_{x}=0$  &
    $u_{0}=v_{0}=w_{0}=\phi_{x}=\psi_{y}=\phi_{y}=\psi_{x}=0$  & \\[20pt]
    \hline
  \end{tabular}
\centering
\end{table}

\begin{figure}[H]
    \centering
    \includegraphics[scale=0.5]{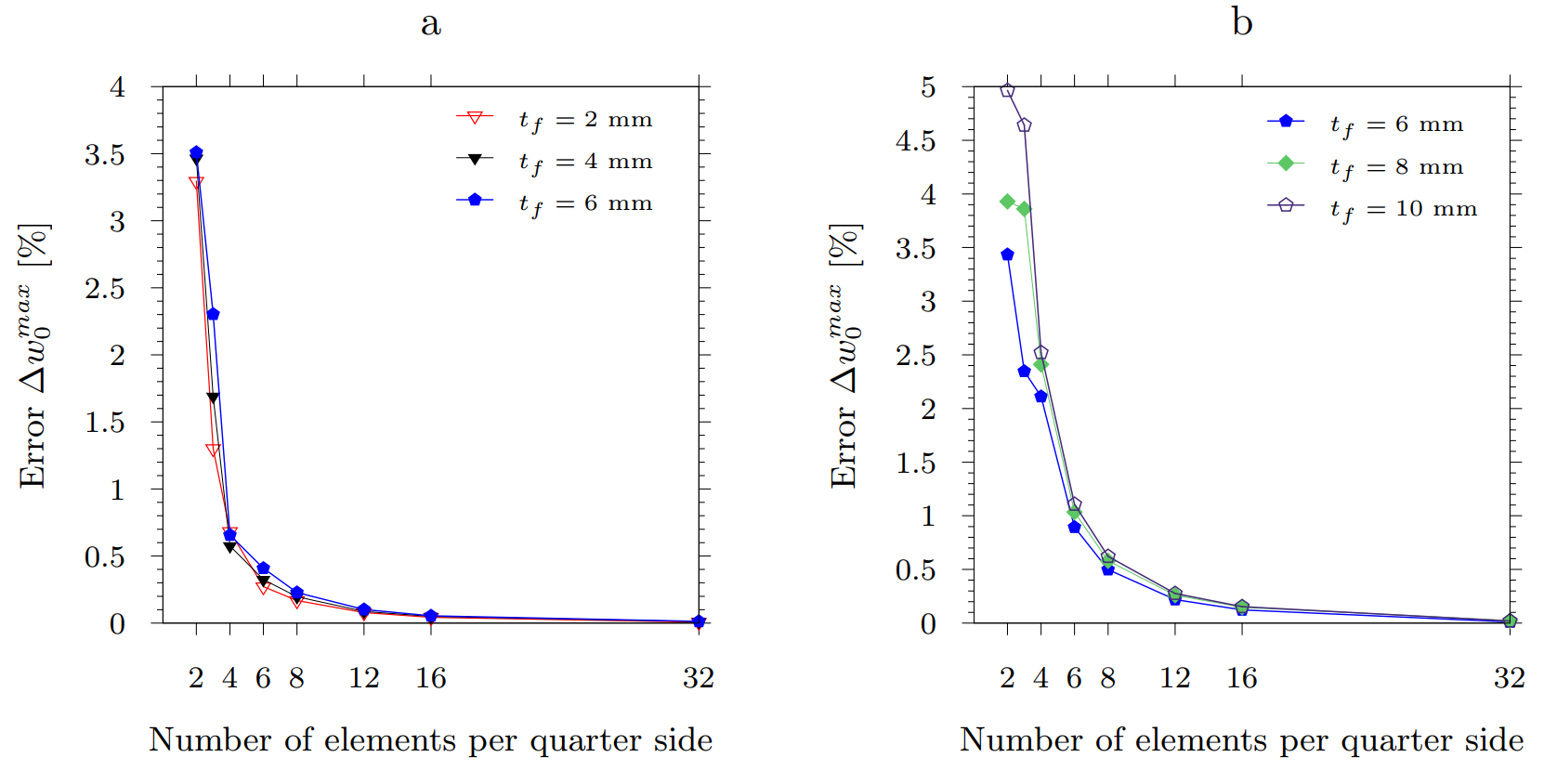}
    \caption{Mesh convergence with respect to maximum transverse deflection $w_{0}^{max}$ for linear analysis on quarter domain of $ (5.4 \cdot 3.6) \textnormal{ m}^{2} $ web-core plates subjected to SSSS boundary conditions. (a) Uniformly distributed load of $ 10000 \textnormal{ N/}\textnormal{m}^2 $  (b) Line load of $ 10000 \textnormal{ N/}\textnormal{m} $ along the $y-$axis.}
    \label{figure 1} 
\end{figure}

A mesh of $32 \times 32$ equal sized rectangular elements on the computational domain was found to give convergent results with respect to the transverse deflection. The mesh convergence results with respect to the linear transverse deflections of web-core lattice plates of size $(5.4 \cdot 3.6) \textnormal{ m}^{2}$ and for various face thickness, $t_{f}$, subjected to SSSS boundary conditions are given in Fig.~{\ref{figure 1}}. The error in maximum transverse deflection is calculated using
\begin{align*}
    \Delta w_{0}^{max} = 100 \times \left(\frac{w_{0}^{max(\textnormal{Nav})}- w_{0}^{max(\textnormal{present FE})}}{w_{0}^{max(\textnormal{Nav})}}\right),
\end{align*}
where $w_{0}^{max(\textnormal{Nav})}$ is the Navier solution to the 2-D micropolar ESL-FSDT plate \citep{karttunen2019}. 

\begin{figure}[H]
    \centering
    \includegraphics[scale=0.48]{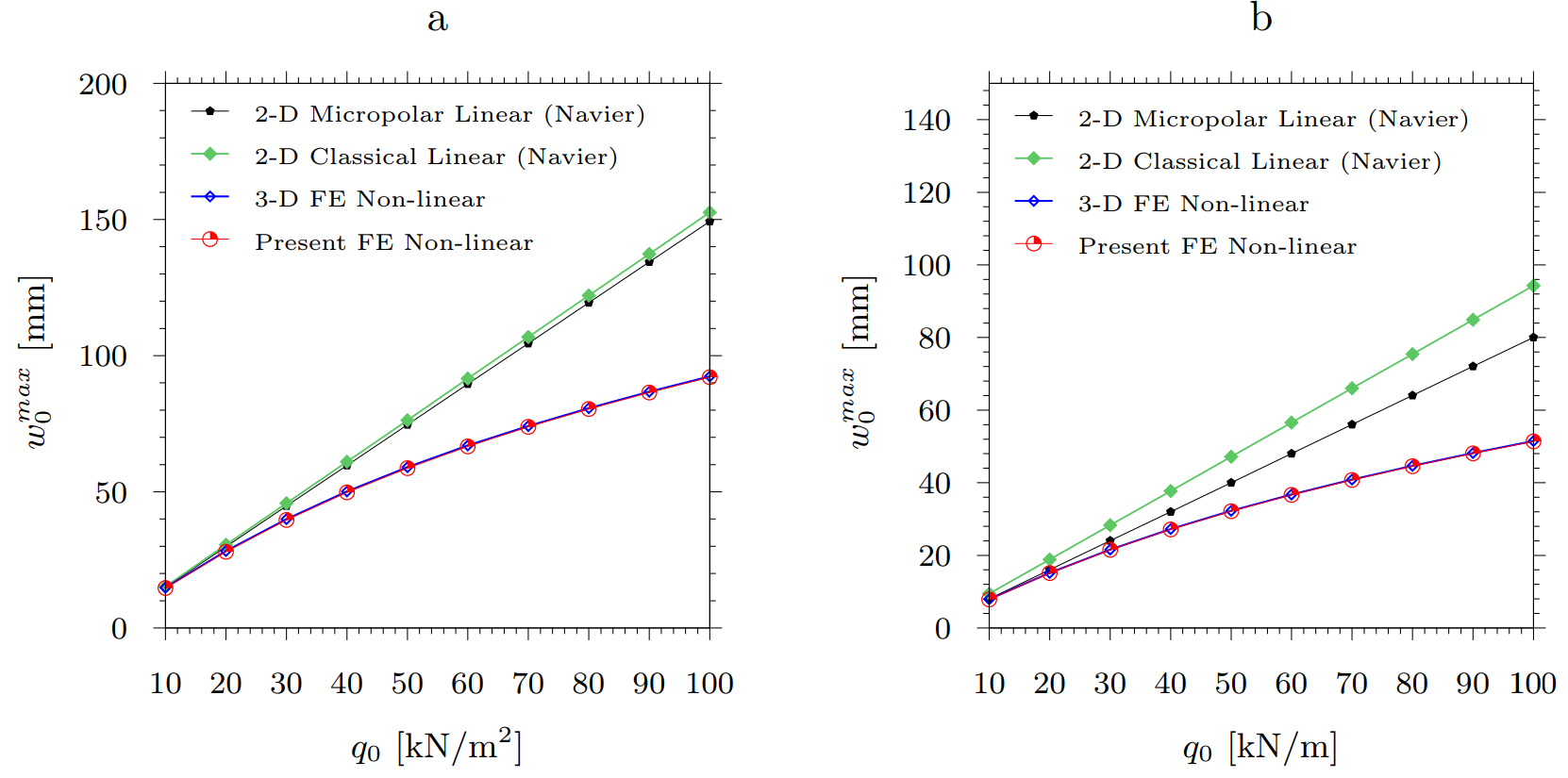}
    \caption{Load vs maximum deflection of $(5.4\cdot 3.6) \textnormal{ m}^{2}$ web-core plate ($t_f = 6 \textnormal{ mm}$) under SSSS boundary conditions. The linear solutions are computed using Navier solution. (a) Uniformly distributed load  (b) Line load along y-axis.}  
    \label{figure 2} 
\end{figure}

Fig.~\ref{figure 2}(a) gives a comparison between the linear transverse deflections of a $(5.4 \cdot 3.6) \textnormal{ m}^{2}$ web-core plate having face thickness $t_{f} = 6 \textnormal{ mm}$, modeled as ESL-FSDT plate based on micropolar elasticity and ESL-FSDT plate based on classical elasticity for uniformly distributed load, while Fig.~\ref{figure 2}(b) shows the comparison for a line load along $y$-axis. The linear transverse deflections are obtained using the Navier solution \citep{karttunen2019,reddy2006}. It can be seen that for the uniformly distributed load the two ESL theories give almost the same transverse deflections but for the line load they deviate from each other. It was shown by \cite{karttunen2019} that the micropolar model predicts the transverse deflections of a line-loaded web-core plate accurately, whereas the classical ESL-FSDT plate yielded displacement errors of 34--175\% for face thicknesses of 2--10 mm. Further, Fig.~\ref{figure 2} also shows the comparison between the nonlinear transverse deflections of the 3-D FE web-core plate (Abaqus) and the corresponding 2-D micropolar plate (present finite element). It can be seen that the present finite element model is able to accurately predict the transverse deflections in both uniformly distributed and line load cases. 

\begin{figure}[hb]
    \centering
    \includegraphics[scale=0.48]{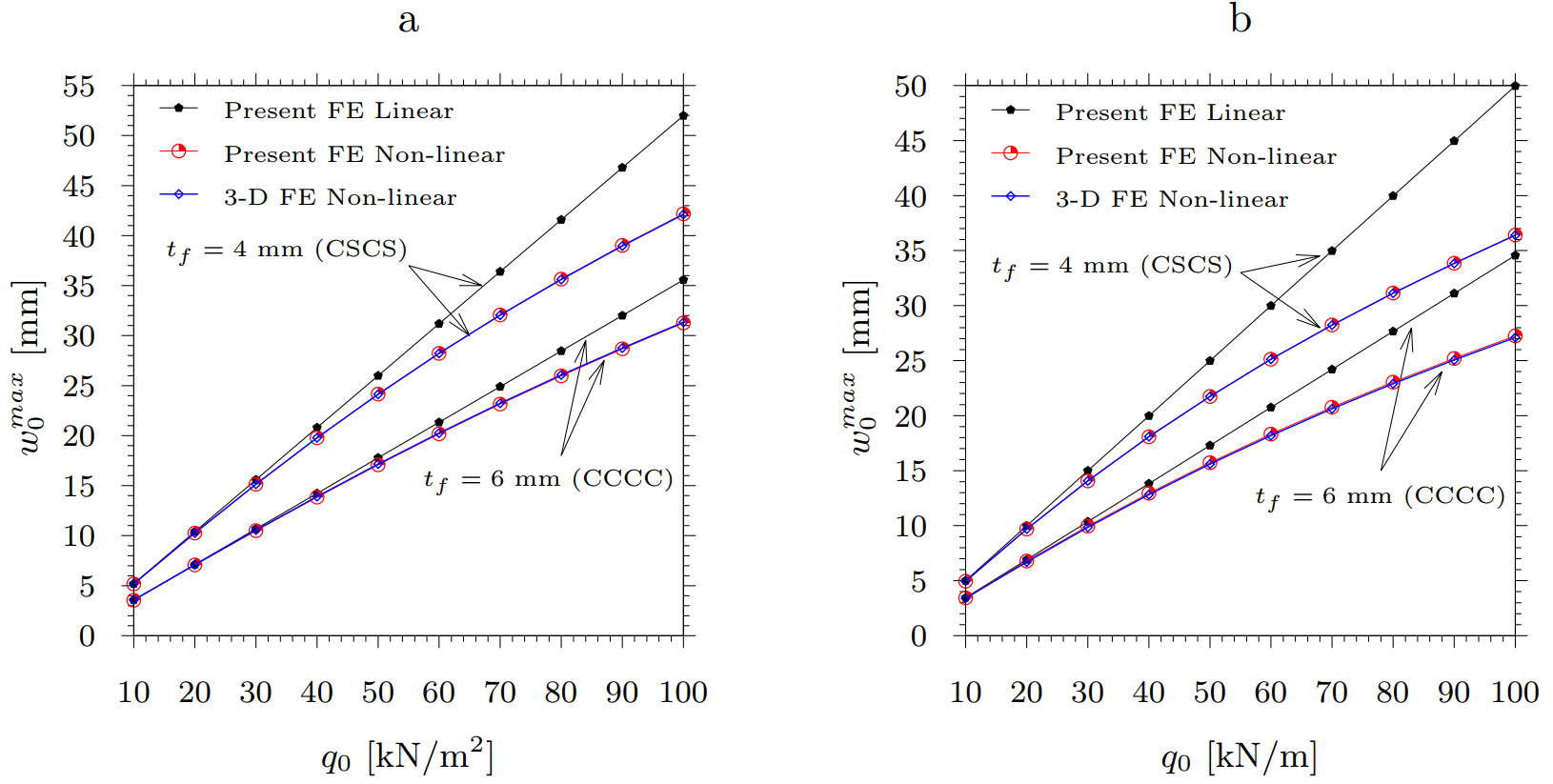}
    \caption{Load vs maximum deflections of $ (5.4 \cdot 3.6) \textnormal{ m}^{2}$ web-core plates with $t_{f} = 4 \textnormal{ mm} $ subjected to CSCS boundary conditions  and $t_{f} = 6 \textnormal{ mm} $ subjected to CCCC boundary conditions. (a) Uniformly distributed load (a) Line load along y-axis.}  
    \label{figure 3} 
\end{figure}

To further test the reliability of micropolar ESL-FSDT plate model and the nonlinear finite element formulation based on it, linear and nonlinear transverse deflections of a $(5.4 \cdot 3.6) \textnormal{ m}^{2}$ web-core plate having face thickness $t_{f} = 4 \textnormal{ mm}$ subjected to CSCS boundary conditions and a $(5.4 \cdot 3.6) \textnormal{ m}^{2}$ web-core plate having face thickness $t_{f} = 6 \textnormal{ mm}$ subjected to CCCC boundary conditions are presented in Fig.~\ref{figure 3} for both uniformly distributed load and line load cases. Moreover, the nonlinear results are compared with the nonlinear results obtained from the 3-D FE analysis of these web-core plates in Abaqus. Excellent agreement between the 2-D micropolar and 3-D reference solutions is observed.

\begin{figure}[H]
    \centering
    \includegraphics[scale=0.5]{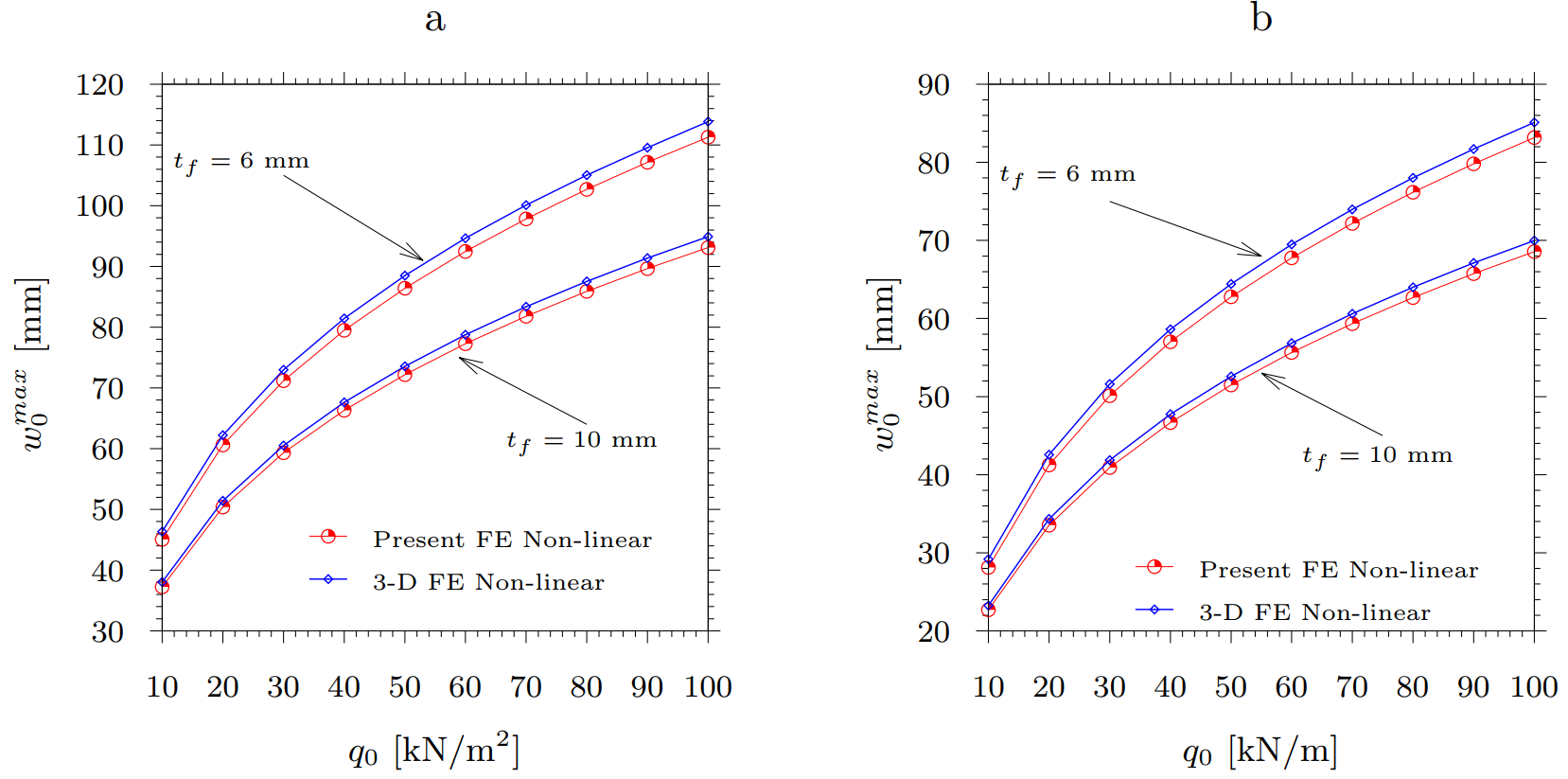}
    \caption{Load vs maximum deflections of $ (5.4 \cdot 3.6) \textnormal{ m}^{2}$ web-core plates with $t_{f} = 6 \textnormal{ mm} $  and $t_{f} = 10 \textnormal{ mm} $ subjected to CFCF boundary conditions. (a) Uniformly distributed load (b) Line load along y-axis.}
    \label{fig:cfcf}
\end{figure}

Finally, we consider the $(5.4 \cdot 3.6) \textnormal{ m}^{2}$ web-core plate with CFCF boundary conditions. In Fig.~\ref{fig:cfcf}(a), web-core plates with face thicknesses $t_{f}=6 \textnormal{ mm}$ and $t_{f}=10 \textnormal{ mm}$ subjected to a uniformly distributed load are considered, while in Fig.~\ref{fig:cfcf}(b) the same plates are under a line load along y-axis (cf.\ Fig.~2). The present 2-D nonlinear finite element model slightly underpredicts the deflections at high load intensities in this case. This maybe due to the fact that Abaqus uses complete Green strain tensor while in the present nonlinear formulation we only considered von K\'arm\'an nonlinear terms. Thus at very high load intensities the von K\'arm\'an nonlinearity may not be an adequate choice for estimating the global deflections. 

It is worth noting that only 1024 isoparametric linear rectangular elements based on Lagrange interpolation functions are used on the computational domain in the 2-D micropolar bending analysis of the $(5.4 \cdot 3.6) \textnormal{ m}^{2}$ web-core lattice plate, while the complete 3-D FE analysis uses 453600 shell elements of type S8R5 as pointed out earlier. Thus, the present finite element model is computationally very efficient in obtaining the global response of lattice plates.  

\subsection{Natural Vibration Frequencies}
Here we will consider the free linear vibration analysis of both web-core and pyramid core lattice plates subjected to the same boundary conditions listed for the bending analysis. But it should be noted that the argument of symmetry cannot be used for the frequency analysis and, thus, the full plate has to be taken as the computational domain. For the linear free vibration analysis using the present finite element model, we ignore the nonlinear terms and consider only the linear terms in evaluating the element coefficient matrices. 

\begin{figure}[H]
    \centering
    \includegraphics[scale=0.5]{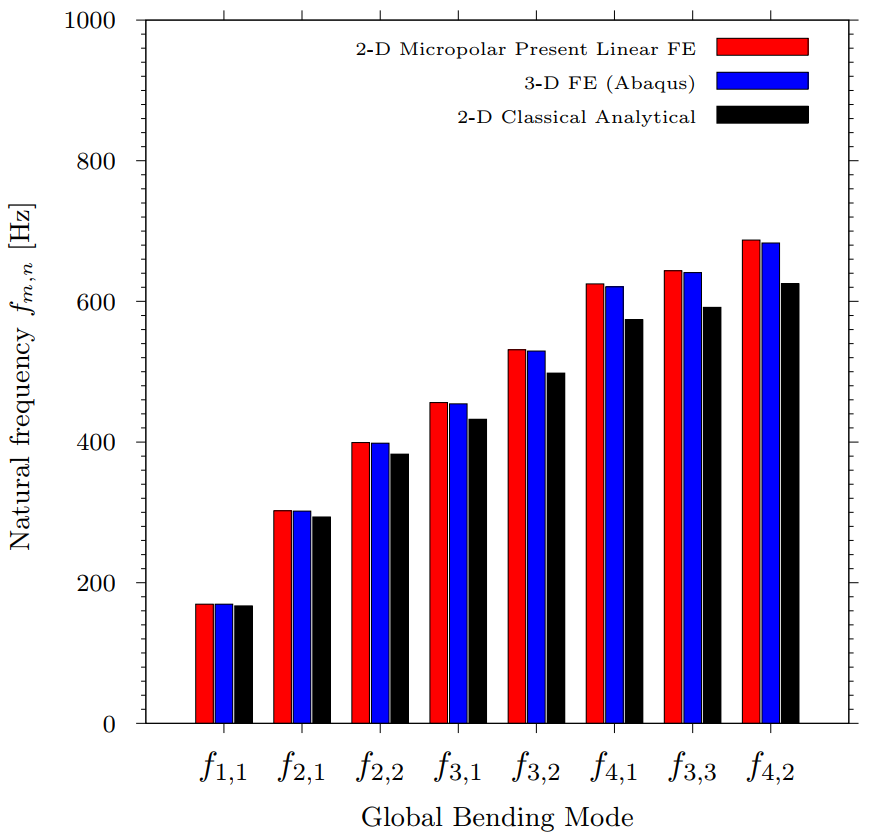}
    \caption{Eight lowest natural vibration frequencies of pyramid core plate of size $(1.0 \cdot 1.0) \textnormal{ m}^{2}$ subjected to SSSS boundary conditions. In $f_{m,n}$, $m$ refers to the number of half waves in $x$-direction and $n$ gives the same for $y$-direction.}
    \label{figure 4}
\end{figure}

In Figure~\ref{figure 4}, a comparison between the 3-D FE analysis and 2-D ESL-FSDT plates based on both classical and micropolar elasticity for the lowest eight natural frequencies of a pyramid core lattice plate of size $(1.0 \cdot 1.0) \textnormal{ m}^{2}$ subjected to SSSS boundary conditions is given. The frequencies from ESL-FSDT micropolar plate model are obtained from the present finite element model while the frequencies of the 2-D classical ESL-FSDT plate are obtained using the Navier solution \citep{karttunen2019}. Both ESL-FSDT plate models provide accurate estimates for the fundamental vibration frequency $f_{1,1}$. However, as the mode number increases, the ESL-FSDT plate based on classical elasticity begins to underpredict the frequencies while the ESL-FSDT plate based on micropolar elasticity still continues to predict the natural frequencies accurately.

\begin{figure}[H]
    \centering
    \includegraphics[scale=0.5]{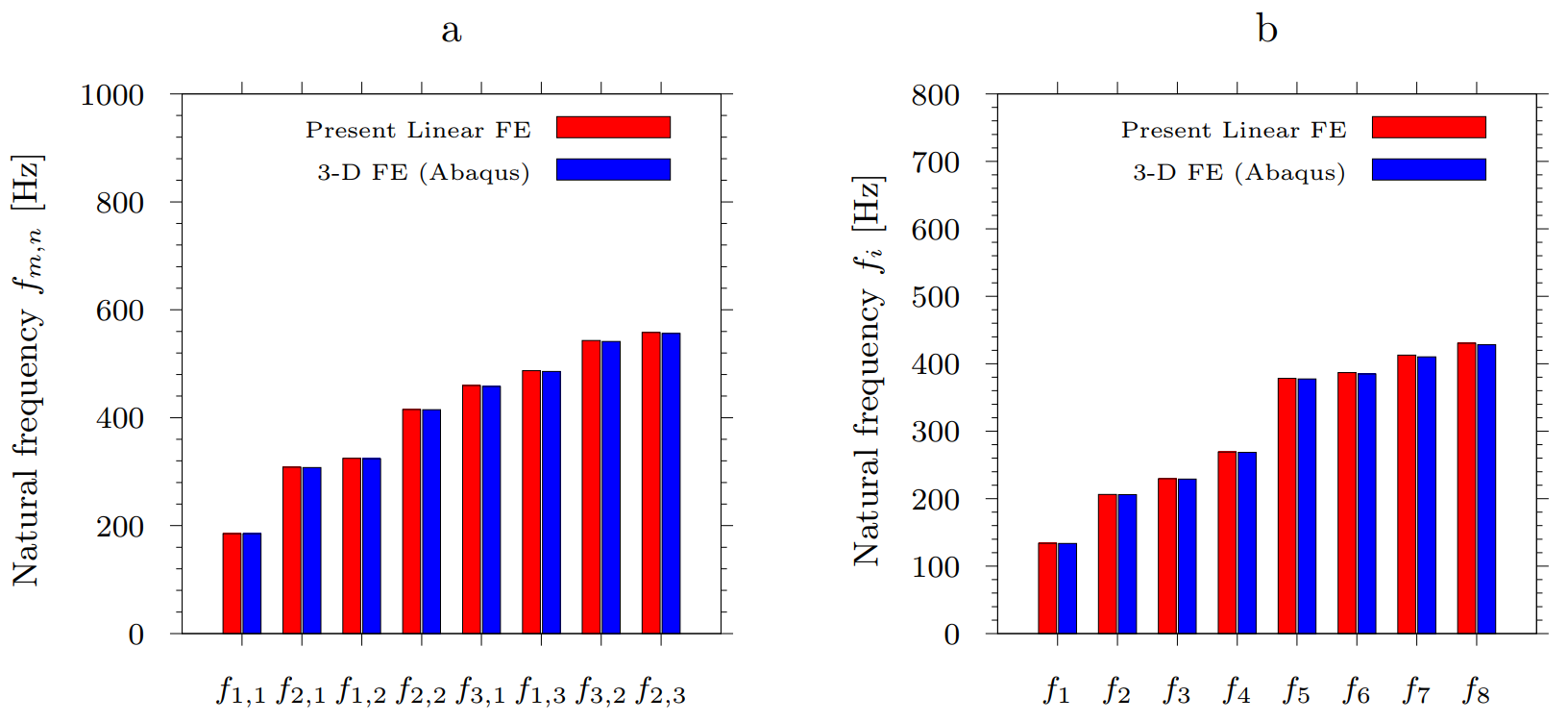}
    \caption{Eight lowest natural vibration frequencies of pyramid core plate of size $(1.0 \cdot 1.0) \textnormal{ m}^{2}$. (a) CSCS boundary condition (b) all edges free (FFFF).}
    \label{figure 5}
\end{figure}

\begin{figure}[H]
    \centering
    \includegraphics[scale=0.5]{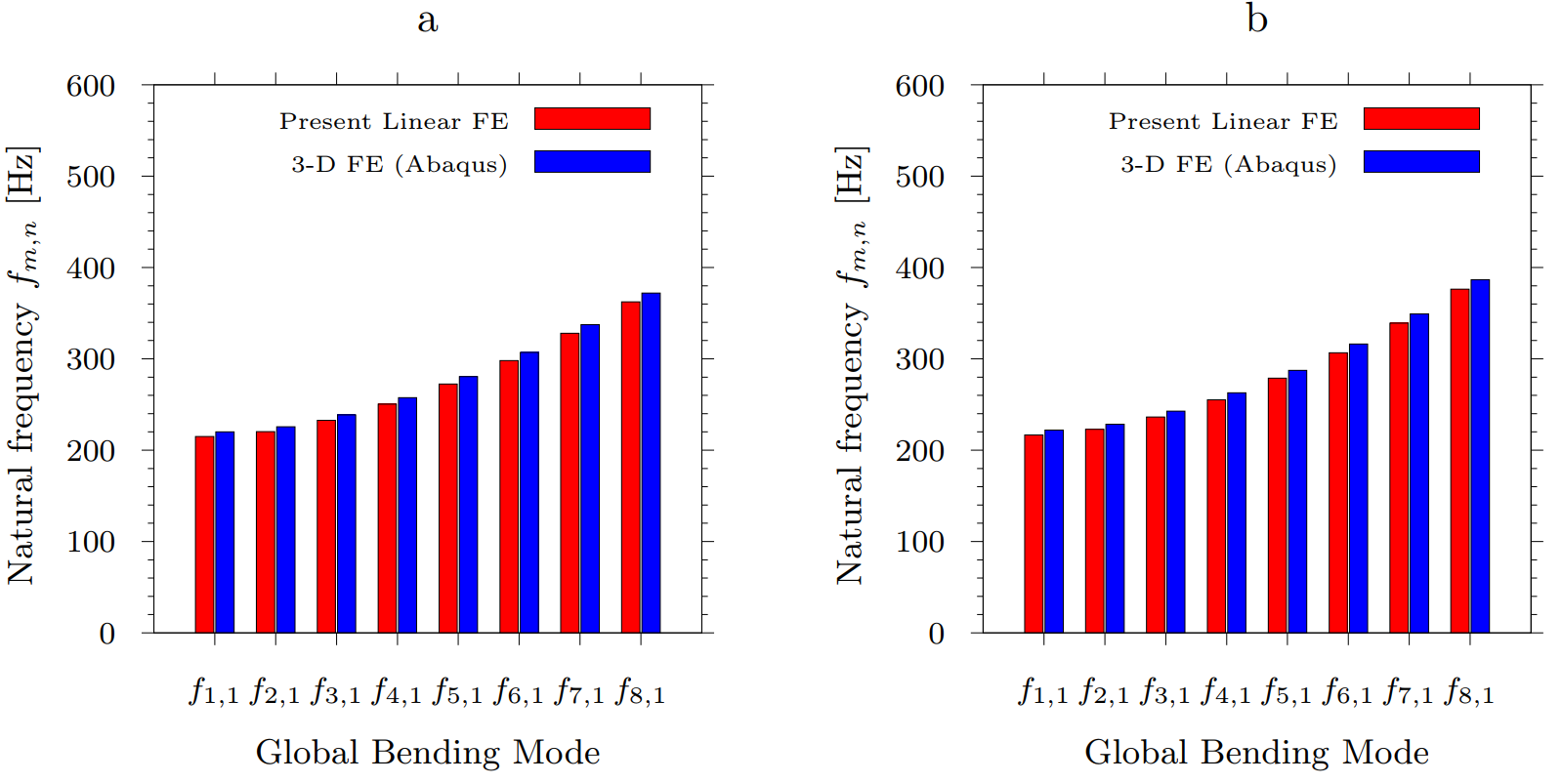}
    \caption{Eight lowest natural vibration frequencies of web-core ($t_{f} = 4 \textnormal{ mm}$)  plate of size $(1.8 \cdot 1.2) \textnormal{ m}^{2}$. (a) CSCS boundary condition (b) CCCC boundary condition.}
    \label{figure 6}
\end{figure}
A comparison between natural frequency results from 3-D FE analysis and from the present finite element formulation of pyramid core lattice plates of size $(1.0 \cdot 1.0) \textnormal{ m}^{2}$ is given in Fig.~\ref{figure 5} for CSCS boundary conditions and an unconstrained plate. Fig.~\ref{figure 6} shows a comparison of 3-D FE results and results from the present finite element formulation for a web-core lattice plate of size $(1.8 \cdot 1.2) \textnormal{ m}^{2}$ having face thickness $t_{f}= 4 \textnormal{ mm}$, subjected to CSCS and CCCC boundary conditions. The 2-D and 3-D results are in good agreement and the 2-D plate provides computationally efficient means for computing the global bending modes of both pyramid core and web-core sandwich panels.
\section{Concluding remarks}
In this paper, a displacement-based geometrically nonlinear finite element model for first-order shear deformation micropolar plates  was developed. Linear Lagrange interpolation functions were used for the generalized displacement variables and appropriate reduced integration techniques were used to overcome shear and membrane locking. The finite element model was used to analyze the bending and free vibrations of lattice core sandwich plates. Specifically, web-core lattice plates with various face thicknesses were considered in bending analysis while both web-core and pyramid core lattice plates were considered in linear vibration analysis. 

Combined with the discussed micropolar constitutive modeling technique for lattice materials, the plate finite element provides efficient means to carry out global bending and vibration analyses for lattice, or architected, core sandwich panels. The micropolar approach is beneficial especially in the case of bending-dominated lattice cores since it considers appropriately both the translations and rotations within 3-D microscale unit cells modeled by classical beam and shell elements when the constitutive modeling for the 2-D macroscale polar plate is carried out. For stretch-dominated cores the micropolar approach becomes relevant especially when a sandwich panel exhibits the thick-face effect \citep{allen1969}. That is to say, when the face sheets are comparatively thick so that their bending and twisting with respect to their own mid-surfaces has to be accounted for by couple stress moments, in addition to the membrane action considered through the usual (classical) moments. Finally, we note that the development of the non-classical micropolar finite element was carried out using standard (classical) techniques, for example, to avoid locking phenomena. The appearance of, for example, both symmetric and antisymmetric shear terms did not essentially lead to any new type of numerical issues. 
\section*{Acknowledgements}
The first and third authors gratefully acknowledge the support of their research through the Oscar S. Wyatt Endowed Chair, while the second author acknowledges the funding received from the European Union's Horizon 2020 research and innovation programme under the Marie Sk\l{}odowska--Curie grant agreement No 745770. The financial support is greatly appreciated. The authors also acknowledge CSC -- IT Center for Science, Finland, for computational resources (Abaqus usage).
\appendix
\section{Finite element matrix components}
The non-zero components of the element stiffness matrix (39) are
\begin{align*}
    K^{11}_{ij} &= \int\limits_{\Omega_{e}}\left(A_{11} \,^{11}S_{ij}^{xx} + A_{44}\,^{11}S_{ij}^{yy} \right)d\Omega, \qquad 
    K^{12}_{ij} = \int\limits_{\Omega_{e}}\left( A_{12}\,^{12}S_{ij}^{xy} + A_{34}\,^{12}S_{ij}^{yx} \right)d\Omega \\
    K^{13}_{ij} &= \int\limits_{\Omega_{e}}\Biggl\{ \frac{A_{11}}{2} \frac{\partial w_{0}}{\partial x} \,^{13}S_{ij}^{xx} + \frac{A_{12}}{2}  \frac{\partial w_{0}}{\partial y} \,^{13}S_{ij}^{xy} + \left(\frac{A_{34}+A_{44}}{4}\right)\left(\frac{\partial w_{0}}{\partial x}\,^{13}S_{ij}^{xy} +\frac{\partial w_{0}}{\partial y}\,^{13}S_{ij}^{yx} \right)\Biggr\}d\Omega \\
    K^{21}_{ij} &= \int\limits_{\Omega_{e}}\left( A_{34}\,^{21}S_{ij}^{xy} + A_{12}\,^{21}S_{ij}^{yx} \right)d\Omega, \qquad  
    K^{22}_{ij} = \int\limits_{\Omega_{e}}\left( A_{33} \,^{22}S_{ij}^{xx} + A_{22} \,^{22}S_{ij}^{yy} \right)d\Omega \\
    K^{23}_{ij} &= \int\limits_{\Omega_{e}}\Biggl\{ \frac{A_{12}}{2} \frac{\partial w_{0}}{\partial x} \,^{23}S_{ij}^{yx} + \frac{A_{22}}{2} \frac{\partial w_{0}}{\partial y} \,^{23}S_{ij}^{yy} + \left(\frac{A_{33}+A_{34}}{4}\right)\left(\frac{\partial w_{0}}{\partial x}\,^{23}S_{ij}^{xy} +\frac{\partial w_{0}}{\partial y}\,^{23}S_{ij}^{xx} \right) \Biggr\}d\Omega \\
    K^{31}_{ij} &= \int\limits_{\Omega_{e}}\Biggl\{ A_{11} \frac{\partial w_{0}}{\partial x} \,^{31}S_{ij}^{xx} + A_{12} \frac{\partial w_{0}}{\partial y} \,^{31}S_{ij}^{yx}   + \left(\frac{A_{34}+A_{44}}{2}\right)\left(\frac{\partial w_{0}}{\partial x}\,^{31}S_{ij}^{yy} +\frac{\partial w_{0}}{\partial y}\,^{31}S_{ij}^{xy} \right)\Biggr\}d\Omega \\
    K^{32}_{ij} &= \int\limits_{\Omega_{e}}\Biggl\{ A_{12}\frac{\partial w_{0}}{\partial x} \,^{32}S_{ij}^{xy} + A_{22}\frac{\partial w_{0}}{\partial y}\,^{32}S_{ij}^{yy} + \left(\frac{A_{33}+A_{34}}{2}\right)\left(\frac{\partial w_{0}}{\partial x}\,^{32}S_{ij}^{yx} +\frac{\partial w_{0}}{\partial y}\,^{32}S_{ij}^{xx} \right) \Biggr\}d\Omega \\
    K^{33}_{ij} & =\int\limits_{\Omega_{e}}\Biggl\{(G_{11}+2G_{12}+G_{22})\,^{33}S_{ij}^{xx} + (G_{33}+2G_{
    34}+G_{44})\,^{33}S_{ij}^{yy}   + \frac{A_{11}}{2}\left(\frac{\partial w_{0}}{\partial x}\right)^{2}\,^{33}S_{ij}^{xx} \nonumber \\
    & + \frac{A_{22}}{2}\left(\frac{\partial w_{0}}{\partial y}\right)^{2}\,^{33}S_{ij}^{yy}  + \frac{A_{12}}{4}\left[\left(\frac{\partial w_{0}}{\partial x}\right)^{2}\,^{33}S_{ij}^{yy} + \left(\frac{\partial w_{0}}{\partial y}\right)^{2}\,^{33}S_{ij}^{xx} + \frac{\partial w_{0}}{\partial y}\frac{\partial w_{0}}{\partial x}\left(\,^{33}S_{ij}^{xy}+\,^{33}S_{ij}^{yx}\right) \right] \nonumber \\
    &  +\left(\frac{A_{33}+A_{34}}{8}\right)\left[\frac{\partial w_{0}}{\partial x}\frac{\partial w_{0}}{\partial y}\left(\,^{33}S_{ij}^{xy}+\,^{33}S_{ij}^{yx}\right) +\left(\frac{\partial w_{0}}{\partial y}\right)^{2}\,^{33}S_{ij}^{xx} + \left(\frac{\partial w_{0}}{\partial x}\right)^{2}\,^{33}S_{ij}^{yy}\right] \nonumber \\
    &+ \left(\frac{A_{34}+A_{44}}{8}\right)\left[\frac{\partial w_{0}}{\partial x}\frac{\partial w_{0}}{\partial y}\left(\,^{33}S_{ij}^{xy}+\,^{33}S_{ij}^{yx}\right) +\left(\frac{\partial w_{0}}{\partial x}\right)^{2}\,^{33}S_{ij}^{yy} +\left(\frac{\partial w_{0}}{\partial y}\right)^{2}\,^{33}S_{ij}^{xx} \right] \Biggr\}d\Omega \\
    K^{34}_{ij} &= \int\limits_{\Omega_{e}}(G_{11}-G_{22})\,^{34}S_{ij}^{x0} d\Omega, \qquad 
    K^{35}_{ij} = 2\int\limits_{\Omega_{e}}(G_{12}+G_{22})\,^{35}S_{ij}^{x0} d\Omega \\
    K^{36}_{ij} &= \int\limits_{\Omega_{e}}(G_{33}-G_{44})\,^{36}S_{ij}^{y0} d\Omega, \qquad
    K^{37}_{ij} = -2\int\limits_{\Omega_{e}}(G_{34}+G_{44}) \,^{37}S_{ij}^{y0} d\Omega \\
    K^{43}_{ij} &= \int\limits_{\Omega_{e}}(G_{11}-G_{22}) \,^{43}S_{ij}^{0x} d\Omega,  \quad
    K^{44}_{ij} = \int\limits_{\Omega_{e}}\left( D_{11} \,^{44}S_{ij}^{xx} + D_{44} \,^{44}S_{ij}^{yy} + (G_{11}-2G_{12}+G_{22})\,^{44}S_{ij}^{00} \right) d\Omega \\
    K^{45}_{ij} &= 2\int\limits_{\Omega_{e}}(G_{12}-G_{22})\,^{44}S_{ij}^{00} d\Omega, \qquad
    K^{46}_{ij} = \int\limits_{\Omega_{e}}\left( D_{12} \,^{45}S_{ij}^{xy} + D_{34} \,^{45}S_{ij}^{yx} \right) d\Omega 
\end{align*}

\begin{align*}
    K^{53}_{ij} &= 2\int\limits_{\Omega_{e}}(G_{12}+G_{22})\,^{53}S_{ij}^{0x} d\Omega  \qquad
    K^{54}_{ij} = 2\int\limits_{\Omega_{e}}(G_{12}-G_{22})\,^{54}S_{ij}^{00} d\Omega \\
    K^{55}_{ij} &= \int\limits_{\Omega_{e}}\left( H_{33} \,^{55}S_{ij}^{xx} + H_{22} \,^{55}S_{ij}^{yy} + 4G_{22} \,^{55}S_{ij}^{00} \right)d\Omega, \qquad 
    K^{57}_{ij} = \int\limits_{\Omega_{e}}\left( H_{34} \,^{57}S_{ij}^{xy} + H_{12} \,^{57}S_{ij}^{yx} \right)d\Omega \\
    K^{63}_{ij} &= \int\limits_{\Omega_{e}}(G_{33}-G_{44}) \,^{63}S_{ij}^{0y} d\Omega, \qquad
    K^{64}_{ij} = \int\limits_{\Omega_{e}}\left( D_{34} \,^{64}S_{ij}^{xy} + D_{12} \,^{64}S_{ij}^{yx} \right) d\Omega \\
    K^{66}_{ij} &= \int\limits_{\Omega_{e}}\left(D_{33} \,^{66}S_{ij}^{xx} + D_{22} \,^{66}S_{ij}^{yy} + (G_{33}-2G_{34}+G_{44}) \,^{66}S_{ij}^{00} \right)d\Omega, \quad
    K^{67}_{ij} = 2\int\limits_{\Omega_{e}}(G_{44}-G_{34}) \,^{67}S_{ij}^{00} d\Omega \\
    K^{73}_{ij} &= -2\int\limits_{\Omega_{e}}(G_{34}+G_{44})  \,^{73}S_{ij}^{0y}  d\Omega, \qquad 
    K^{75}_{ij} = \int\limits_{\Omega_{e}}\left( H_{12}  \,^{75}S_{ij}^{xy}  + H_{34}\frac{\partial L^{(7)}_{i}}{\partial y}  \,^{75}S_{ij}^{yx}  \right) d\Omega \\
    K^{76}_{ij} &= 2\int\limits_{\Omega_{e}}(G_{44}-G_{34}) \,^{76}S_{ij}^{00} d\Omega,   \qquad
    K^{77}_{ij} = \int\limits_{\Omega_{e}}\left(H_{11} \,^{77}S_{ij}^{xx} + H_{44} \,^{77}S_{ij}^{yy} + 4G_{44} \,^{77}S_{ij}^{00} \right)d\Omega 
\end{align*}
Simlarly the non-zero components of element mass matrix (38) are given by

\begin{align*}
    M^{11}_{ij} &= \int\limits_{\Omega_{e}}m_{11} \,^{11}S_{ij}^{00} d\Omega, \quad M^{22}_{ij} = \int\limits_{\Omega_{e}}m_{22} \,^{22}S_{ij}^{00} d\Omega, \quad M^{33}_{ij} = \int\limits_{\Omega_{e}}m_{33} \,^{33}S_{ij}^{00} d\Omega, \quad  M^{44}_{ij} = \int\limits_{\Omega_{e}}m_{44} \,^{44}S_{ij}^{00} d\Omega\\
    M^{45}_{ij} &= \int\limits_{\Omega_{e}}m_{45} \,^{45}S_{ij}^{00} d\Omega, \quad M^{54}_{ij} = \int\limits_{\Omega_{e}}m_{45} \,^{54}S_{ij}^{00} d\Omega, \quad  M^{55}_{ij} = \int\limits_{\Omega_{e}}m_{55} \,^{55}S_{ij}^{00} d\Omega, \quad M^{66}_{ij} = \int\limits_{\Omega_{e}}m_{66} \,^{66}S_{ij}^{00} d\Omega \\
    M^{67}_{ij} &= \int\limits_{\Omega_{e}}m_{67} \,^{67}S_{ij}^{00} d\Omega, \quad   M^{76}_{ij} = \int\limits_{\Omega_{e}}m_{67} \,^{76}S_{ij}^{00} d\Omega, \quad M^{77}_{ij} = \int\limits_{\Omega_{e}}m_{77} \,^{77}S_{ij}^{00} d\Omega 
\end{align*}
where we have used the notation
\begin{align*}
    ^{IJ}S_{ij}^{ab} = \frac{\partial L_{i}^{(I)}}{\partial a} \frac{\partial L_{j}^{(J)}}{\partial b},\quad ^{IJ}S_{ij}^{0b} = L_{i}^{(I)} \frac{\partial L_{j}^{(J)}}{\partial b}, \quad ^{IJ}S_{ij}^{a0} = \frac{\partial L_{i}^{(I)}}{\partial a} L_{j}^{(J)}, \quad ^{IJ}S_{ij}^{00} = L_{i}^{(I)} L_{j}^{(J)} 
\end{align*}
where $I,J = \{1,2,3,4,5,6,7\}$, $i,j = \{1,2,3,4\}$ and $a,b=\{x,y\}$.

Similarly, the components of the element tangent stiffness matrix are given 
by
\begin{align*}
    T_{ij}^{IJ} = K_{ij}^{IJ}
\end{align*}
except for the following terms
\begin{align*}
    T^{13}_{ij} &= 2K^{13}_{ij}, \quad T^{23}_{ij} = 2K^{23}_{ij} \\
\end{align*}
\begin{align*}
    T^{33}_{ij} &= K^{33}_{ij} + \int\limits_{\Omega_{e}}\Biggl\{ A_{11}\frac{\partial u_{0}}{\partial x}\,^{33}S_{ij}^{xx}  + A_{12}\frac{\partial u_{0}}{\partial x}\,^{33}S_{ij}^{yy} + \left(\frac{A_{34}+A_{44}}{2}\right)\left(\frac{\partial u_{0}}{\partial y}\,^{33}S_{ij}^{xy}  +\frac{\partial u_{0}}{\partial y}\,^{33}S_{ij}^{yx}  \right) \Biggr\} d\Omega  \nonumber \\
    & \quad + \int\limits_{\Omega_{e}}\Biggl\{ A_{12}\frac{\partial v_{0}}{\partial y}\,^{33}S_{ij}^{xx}  + A_{22}\frac{\partial v_{0}}{\partial y}\,^{33}S_{ij}^{yy}  + \left(\frac{A_{33}+A_{34}}{2}\right)\left(\frac{\partial v_{0}}{\partial x}\,^{33}S_{ij}^{xy}  +\frac{\partial v_{0}}{\partial x}\,^{33}S_{ij}^{yx}  \right) \Biggr\}d\Omega  \nonumber \\
    & \quad + \int\limits_{\Omega_{e}}\Biggl\{ A_{11}\left(\frac{\partial w_{0}}{\partial x}\right)^{2}\,^{33}S_{ij}^{xx} + A_{22}\left(\frac{\partial w_{0}}{\partial y}\right)^{2}\,^{33}S_{ij}^{yy}  \nonumber \\
    & \quad  + \left(\frac{A_{12}+A_{34}}{4}+\frac{A_{33}+A_{44}}{8}\right)\left[\left(\frac{\partial w_{0}}{\partial x}\right)^{2}\,^{33}S_{ij}^{yy} + \left(\frac{\partial w_{0}}{\partial y}\right)^{2}\,^{33}S_{ij}^{xx} \right] \nonumber \\
    & \quad + \left(\frac{A_{12}+A_{34}}{4}+\frac{A_{33}+A_{44}}{8}\right)\left[3\frac{\partial w_{0}}{\partial y}\frac{\partial w_{0}}{\partial x}\left(\,^{33}S_{ij}^{xy} + \,^{33}S_{ij}^{yx}\right)\right] \Biggr\}d\Omega
\end{align*}  
\section{Pyramid core parameter values}
The constitutive parameters of the pyramid core presented in Fig.~1 are given in Table B.1 and the inertia coefficients in Table B.2. The parameter values for the web-core unit cell were given in \cite{karttunen2019}.
\setcounter{table}{0}
\begin{table}[H]
\centering
  \caption{Constitutive parameters for the pyramid core presented in Fig.~1}
  \begin{tabular}{M{0.4in}|M{0.8in}|M{0.4in}|M{0.8in}|M{0.4in}|M{0.8in}|M{0.4in}|M{0.8in} @{}m{0pt}@{}}
    \hline
     &
     $\mathbf{A} \textnormal{ [MN/m]}$ &
      &
     $\mathbf{D} \textnormal{ [kN/m]}$ & 
      &
     $\mathbf{G} \textnormal{ [MN/m]}$ &
     &
     $\mathbf{H} \textnormal{ [Nm]}$  \\
    \hline
    $A_{11}$ & 943.83 & $D_{11}$ & 294.96  & $G_{11}$ & 11.907 & $H_{11}$ & 1603.5 & \\
    $A_{12}$ & 312.24 & $D_{12}$ & 97.579  & $G_{12}$ & 9.1773 & $H_{12}$ & 18.615 & \\
    $A_{22}$ & 943.83 & $D_{22}$ & 294.96  & $G_{22}$ & 9.1770 & $H_{22}$ & 1603.5 & \\
    $A_{33}$ & 317.19 & $D_{33}$ & 99.126  & $G_{33}$ & 11.907 & $H_{33}$ & 2829.5 & \\
    $A_{34}$ & 317.19 & $D_{34}$ & 99.126  & $G_{34}$ & 9.1773 & $H_{34}$ & -933.59 & \\
    $A_{44}$ & 317.19 & $D_{44}$ & 99.126  & $G_{44}$ & 9.1770 & $H_{44}$ & 2829.5 & \\
    \hline
  \end{tabular}
\centering
\end{table}
\begin{table}[H]
\caption{Inertia coefficients for the pyramid core presented in Fig.~1}
\centering
  \begin{tabular}{M{0.4in}|M{0.8in}|M{0.4in}|M{0.8in} @{}m{0pt}@{}}
    \hline
     &
     $\mathbf{M}\textnormal{ [kg}/\textnormal{m}^{2}]$ &
      &  
      $\mathbf{M} \textnormal{ [kg}]$  \\
    \hline
    $m_{11}$ & 33.264 & $ m_{44}$ & 0.0102 & \\
    $m_{22}$ & 33.264 & $ m_{45}$ & 0 & \\
    $m_{33}$ & 33.264 & $ m_{55}$ & 0.1715$\cdot10^{-3}$ & \\
     & & $ m_{66}$ & 0.0102 & \\
     & & $ m_{67}$ & 0 & \\
     & & $ m_{77}$ & 0.1715$\cdot10^{-3}$ & \\
    \hline
  \end{tabular}
\centering
\end{table}

\bibliographystyle{elsarticle-harv}
\bibliography{microliteraFE}






\end{document}